\documentclass[useAMS,usenatbib,twocolumn]{mn2e}

\usepackage{graphicx}
\usepackage{amsmath}
\usepackage{amssymb}
\usepackage{float}
\usepackage[draft]{hyperref}

\usepackage{color}

\newcommand{\ha}{H$\alpha$}
\usepackage{color}

\begin{document}

\title[Astrophysical Tests of Modified Gravity]{Astrophysical Tests
of Modified Gravity: the Morphology and Kinematics of Dwarf Galaxies} 
\author[Vikram, V. {\it et al.}]{Vinu Vikram, Anna Cabr\'e, Bhuvnesh Jain, Jake  
VanderPlas\\
  Department of Physics and Astronomy, University of Pennsylvania, 
  Philadelphia, PA 19104\\
  Center for Particle Cosmology, University of Pennsylvania, 
  Philadelphia, PA 19104\\
  Astronomy Department, University of Washington,
  Seattle, WA 98195\\
  Department of Computer Science and Engineering, University of Washington,
  Seattle, WA 98195
}

\date{\today}
\maketitle

\begin{abstract}
This paper is the third in a series on tests of gravity using observations of 
stars and nearby
dwarf galaxies. We carry out four distinct tests using published data on the
kinematics
and morphology of dwarf galaxies, motivated by the theoretical work of
Hui et al. (2009) and Jain \& Vanderplas (2011). In a wide class of
gravity theories a scalar field couples to matter and provides an attractive
fifth force. Due to their different self-gravity,
stars and gas may respond differently to the scalar force leading
to several observable deviations from standard gravity.  
HI gas, red giant stars and main sequence stars
can be displaced relative to each other,
and the stellar disk can display warps or asymmetric rotation curves
aligned with external potential gradients.  
To distinguish the effects of modified
gravity from standard astrophysical phenomena, we use a control sample of
galaxies that are expected to be screened from the fifth force.
In all cases we find no significant deviation
from the null hypothesis of general relativity. The limits obtained from dwarf galaxies are not yet competitive with the limits from cepheids obtained in our first paper, but
can be improved to probe regions of parameter space that are inaccessible using
other tests. We discuss how our methodology can be applied to 
new radio and optical observations of nearby galaxies.
\nocite{hui09, bhuvjake2011}
\end{abstract}


\section{Introduction}
The explanation of the observed accelerated expansion of the universe remains a
mystery, but possible solutions to this puzzle fall into two broad
categories. The first is to posit the existence of a new component of the
universe with an appropriate equation of state to cause the observed
acceleration. This ``dark energy'' may be due to an exotic particle or field, or
be related to the vacuum energy of space itself. A second approach is to seek to
explain this acceleration through modifications to the field equations of
general relativity (GR) itself. In recent years this modified gravity (MG)
approach has received attention and various approaches have been actively
developed -- for a review see e.g. \citet{jain-khoury10}. 
Even on theoretical grounds, modified theories of gravity are well
motivated: GR is unlikely to be the complete theory of gravity owing to 
its singularities and non-renormalizeability. Hence it is generally
considered to be the low energy effective action of some UV-complete
theory, though our interest is modifications in the long distance/low energy
regime. 
 
 A modification of GR on large (astrophysical) scales generically leads to
scalar-tensor theories of gravity, where a new scalar field couples to gravity.
Equivalently these theories can be described via a coupling of the scalar field
to matter, which leads to enhancements of the gravitational force. 
Nonrelativistic matter -- such as the stars, gas, and dust in galaxies -- will
feel this enhanced force, which in general leads to larger dynamically inferred
masses. The discrepancy can be up to a factor of 1/3 in f(R) or DGP gravity. We
note that photons, being relativistic, do not feel the enhanced force, so that
lensing measurements probe the true mass distribution.  
 
This enhanced gravitational force should be detectable through fifth force 
experiments, tests of the equivalence principle (if the scalar coupling to
matter varied with the properties of matter) or through the orbits of planets
around the Sun \citep{will06}. \citet{khoury04} proposed that
nonlinear screening of the scalar field, called chameleon screening, can
suppress the fifth force in high density environments such as the Milky Way, so
that Solar System and lab tests can then be satisfied. This screening was
originally suggested to hide the effects of a quintessence-like scalar that
forms the dark energy and may couple to matter -- generically such a coupling is
expected unless forbidden by a symmetry -- so there are theoretical
reasons to expect 
a screening effect to operate in either a dark energy or modified gravity
scenario. Indeed, since there are only a handful of screening mechanisms
(Vainshtein and symmetron screening are discussed below), small scale tests of
gravity that rely on distinct signatures of screening are useful discriminators
of cosmological models.

In this paper we will consider deviations from GR exhibited in
theories that rely on chameleon screening.
Qualitatively similar behavior occurs in symmetron screening
\citep{hinterbichler10} and the environmentally dependent
dilaton \citep{brax10} and the tests we  present here apply
to these mechanisms as well\footnote{In order to identify the screened and
  unscreened galaxies it is necessary to calibrate on N-body simulations,
  which give different criteria for different screening mechanisms. While
  we do not implement criteria for any mechanism other than chameleon,  
  our results appear robust to the details of screening.}. 

While local dynamical studies suggest that any fifth-force effects must be
screened within the Milky Way, 
dwarf galaxies in low-density environments may remain unscreened as the 
Newtonian potential (which determines the level of screening) can be an order of
magnitude 
smaller in magnitude than in the Milky Way. Hence dwarf galaxies can
exhibit manifestations of modified forces in  their infall motions, 
internal dynamics and segregation of stars, black holes and gas. \citet{hui09}, \citet{hui10}, \citet{bhuvjake2011},
\citet{changhui}, \citet{Davis2012} and 
\citet{jainvinu2012}
discuss various observational consequences. Indeed the small scale dynamics of 
nearby galaxies may have a bigger signal of MG than large-scale perturbations 
\citep[see][for a discussion of observational approaches]{jain11}.
 
In this work, we discuss observable effects on disk galaxies that  arise due to 
their internal dynamics or their interaction with a neighbor or other sources of
an external gravitational field. We focus on late-type dwarf galaxies as these
are most likely to be unscreened. The observational effects are rooted in the
fact that compact objects like stars within
the dwarf galaxies can self-screen, and thus respond to a different
gravitational potential than the diffuse unscreened components like gas and dark matter.
So, for example, if the stellar component of a dwarf galaxy self-screens,
it will lag the dark matter and HI gas disk in the infall towards
another galaxy. This may lead to a separation of the stellar disk from the
center of mass of the dark matter and from the HI disk, and lead to
observable distortions of the morphology and/or dynamics of the stellar disk.
\citet{bhuvjake2011} computed a variety of dynamical consequences of these
effects. 
 
However, the origin of modified gravity effects
may be difficult to disentangle from other astrophysical
processes. To address this, we create a control samples of screened galaxies
which are not expected to show any of the expected modified gravity effects.
By comparing statistics of the unscreened sample and screened control
sample, we may derive constraints on modified gravity parameters.
The division of observed galaxies into screened and unscreened samples
is accomplished based on an estimate of
the local value of the external gravitational potential at the location
of each galaxy: see \cite{cabre2012} for a description of the methodology
used to distinguish the screened and unscreened samples.
\cite{jainvinu2012} used this methodology and
obtained modified gravity constraints
based on cepheid variable stars and other distance indicators.
This paper is the third in the series of papers on astrophysical tests of gravity: 

In this paper we perform several of the proposed tests of modified gravity
using currently available data. Because of the low signal-to-noise of these
types of tests, a large amount of high quality data will be required to
obtain competitive constraints on modified gravity models.
Therefore, one of the main goals of this paper is to show the strengths and
weaknesses of currently available data for these tests, and explore the
requirements for future data in this area.

This paper is organized as follows. In section \ref{pot-sec} we describe the
method to generate potential map of the local Universe, which allows us to
prepare our samples of screened and unscreened galaxies.
In section  \ref{sec:test-summary} we briefly summarize each of the observable
effects we will use to test modified gravity.
In section \ref{sec:offset} we describe the offset between stellar
and gaseous components, which includes:
offset between HI and optical centroids (section \ref{sec:h1-opt-offset}), 
offset between optical center and \ha{} center (section \ref{sec:kinematical}), 
and offset between red giants and main sequence stars  (section
\ref{sec:rgb}).
In section \ref{sec:warp} we study MG-induced warping of edge-on galaxies.
In section \ref{sec:rotationcurves} we describe the tests based on
 gaseous velocity rotation curves, which includes
HI - \ha{} velocity differences (section \ref{sec:hihalpha}), and
asymmetry in the \ha{} curves (section \ref{sec:asymmetry}).
We conclude in section \ref{sec:discussion}.

\section{Potential map of the local Universe}
\label{pot-sec}
Each of the modified gravity tests presented in this paper relies on comparing
various observationally-derived statistics of a screened and unscreened sample
of galaxies.
Therefore, the first step of the analysis is to determine whether each
galaxy in our sample is likely to be screened, i.e. the `fifth' force due
to the  scalar field is suppressed by the deep 
potential well of the galaxy or its environment.  
Because the screened sample should be free of any fifth-force effects, we use
it as a control sample.  The detailed procedure for 
determining screening has been described and tested via N-body simulations
of $f(R)$ gravity in \cite{cabre2012}; we summarize these results here.

For an isolated spherical halo in $f(R)$ gravity,
the chameleon effect is suppressed if $f_{R0} \leq 2/3\
|\phi_N|/c^2$, where $\phi_N$ is the Newtonian  potential of the object
\citep{hu07}. We will approximate galaxies as isolated spherical halos,
and make use of this condition to determine whether an
object is self-screened or environmentally screened: i.e. we look for the
galaxies which satisfy the following conditions: 
\begin{equation}
\label{eq:criterion} 
 \frac{|\phi_{\rm int}|}{c^2}>\frac{3}{2} f_{R0}\; \; \; \; \; \; \; \mathrm{or}
\; \; \; \; \; \; \; \frac{|\phi_{\rm ext}|}{c^2}>\frac{3}{2} f_{R0}
\end{equation}
The first condition describes the {\it self-screening} condition, with 
$ |\phi_{\rm int}|= GM/r_{\rm vir}$, where the $M$ and $r_{\rm vir}$ are the
mass and virial radius. We typically use measurements of the circular velocity at large radii to establish self-screening, since we work with rotating disk galaxies for our tests. 
The second condition describes {\it environmental
screening} condition, with $|\phi_{\rm ext}|$ representing the
Newtonian potential due to the neighbor galaxies within the background Compton
length $\lambda_C$.
The external potential $|\phi_{\rm ext}|$ is evaluated using 
neighboring galaxies as:
\begin{equation}\label{eq:criterion22} 
 |\phi_{\rm ext}|= \sum_{d_i<\lambda_C+r_i} \frac{GM_i}{d_i}
\end{equation}
where $d_i$ is the physical distance to the neighboring
galaxy with mass $M_i$ and virial radius $r_i$. $\lambda_C$ is related
to the background field value $f_{R0}$
\citep{sch09}:
\begin{equation}
 \lambda_C \approx 32 \sqrt{\frac{\left|f_{R0}\right|}{10^{-4}}} \mathrm{ Mpc}
 \label{eq:compton}
\end{equation}

Identifying screened and unscreened regions requires knowledge of the masses and
three-dimensional positions of all the objects in the sky.
Since our knowledge is limited by finite survey sizes and depths,
we assume that most of the contribution to the environmental screening
potential comes from
clusters and groups of galaxies. This is a reasonable assumption as the
potential of a rich cluster is up to 
two orders of magnitudes larger than a typical galaxy. \citet{cabre2012}
compiled a catalog of bright clusters with known redshifts based on 
Abell \citep{abe89} and ROSAT \citep{ebe96} observations. We remove many
clusters from the Abell catalog as they lack spectroscopic
redshift or have redshift greater than 0.1. Our final catalog contains 675
clusters of galaxies with galactic latitude $\left|b\right| > 20$.

The next most significant contribution comes from galaxy groups.
\citet{yan07} provides a nearly complete catalog of groups in the SDSS region,
including around 400,000 galaxies
covering the redshift range $0.01-0.2$.
In addition, we use the group catalog from the 2M++ all
sky survey \citep{lav011} to create a combined catalog of
153380 SDSS and 3984 2M++ galaxy groups.
The final contribution to the catalog are
local galaxies within 10 Mpc, identified by \citet{kar04}. 

With this catalog in place, an estimate of the mass of each object
is required in order to calculate the potential.
We determine the mass using either the mass-luminosity relation given by
\citet{rei02} or the mass-velocity dispersion relation given by \citet{evr08}.
The masses of SDSS groups and local galaxies are taken directly from
\citet{yan07} and \citet{kar04}, respectively\footnote{Note
that \citet{kar04} reports the mass for only 313 of the 451 galaxy sample.
See \href{http://www.sas.upenn.edu/~vinu/screening/index.html}{
http://www.sas.upenn.edu/$\sim$vinu/screening/index.html} for an extensive
description of the catalog.}.

Using the mass of the clusters, groups and galaxies, we generate a
potential map of the sky.
In this paper we use values of the Compton length between $\lambda_C
= 3.2$ Mpc and  $\lambda_C = 1.0$ Mpc corresponding to $f_{R0}$ parameters
$10^{-6}$ and $10^{-7}$.  
Due to survey incompleteness, the derived potential map
is only an approximation for many regions of the sky. This could potentially
lead to systematic inaccuracies in labeling of screened and unscreened objects.
A detailed discussion of these uncertainties can be found in \citet{cabre2012}. 
 
\section{Observable Effects Modified Gravity}
\label{sec:test-summary}

In chameleon gravity theories, objects with large gravitational potential
may be self-screened (see Eq.~\ref{eq:criterion}).  This may lead to observable
differences in the dynamics of different galactic components:  
the gaseous disk, dark matter halo, and giant
stars may feel the fifth force, while the self-screened main sequence
stars move according to standard gravitational dynamics.
The different forces on these components lead to several potentially observable
morphological and kinematical differences between the two components. In
this section, we summarize the expected effects of the fifth force, as well
as potential observational tracers of these effects.

\subsection{Summary of physical effects of modified gravity}
The following effects of modified gravity are expected to appear in 
unscreened dwarf galaxies.
Quantitative estimates of the magnitude of these
observables are given in the following sections.
For more details and motivation of these predictions, see \citet{bhuvjake2011}. 

\begin{enumerate}
\item \textbf{Offset between stellar and gaseous components (section
\ref{sec:offset}):}
Chameleon theories predict that in the presence of a suitable
potential gradient, low density, unscreened components like dark matter or
neutral hydrogen clouds (HI) will fall faster along an external potential
than will self-screened stars. 
This difference in acceleration may cause a measurable offset between the
centroid of the stellar disk and the gaseous disk. 
The population of red giant stars, being partially screened, may also show
an observable offset from the population of main sequence stars.
The size of the offset
depends strongly on the mass and concentration of the dwarf galaxy's halo,
as well as the distance to neighboring galaxies.
\citet{bhuvjake2011} estimate this offset may be several percent of the virial
radius of suitable dwarf galaxies. We carry out three tests of such offsets in 
section 4.  

\item \textbf{Warping of the stellar disk (section \ref{sec:warp}):}
As the halo moves along an external potential gradient,
it pulls at the lagging stellar component.  If the potential gradient
is aligned with the axis of rotation, the steep slope of the halo potential
may introduce a U-shaped warp in the stellar disk.
The shape and strength of the warp depends on the mass profile of the dwarf
galaxy halo, as well as the strength of the external potential gradient. The
warp  
 is expected to align with this potential gradient.
\citet{bhuvjake2011} estimate the warp to be of order $0.1$ kpc, though it
varies significantly depending on the external fifth force acting on the dwarf
galaxy.  

\item \textbf{Rotation curve tests (section \ref{sec:rotationcurves}):}
If a disk galaxy falls edge-on, or more generally if the external potential
gradient is not aligned with the galaxy's axis of
rotation, the offset between stellar and halo centers can perturb the stellar
disk
and cause an asymmetry in the stellar rotation curve. This is due to the fact
that the dominant force on the stellar disk is from the potential of the halo,
rather than the potential of the disk itself. The asymmetry
can be as high as 10 km/s. 
The asymmetry can be distinguished from standard astrophysical sources  
by comparing the stellar disk with the gaseous disk.  A version of this test is
show in section 6.2. 

In principle the rotation velocity of an unscreened gaseous
disk is also enhanced relative to the stellar disk by a factor of
$\sqrt{1 + \Delta G/G}$, where $\Delta G$ is the contribution of the fifth force
to the self-gravity of a galaxy \citep{hui09}. 
We find that current data
is unsuitable to carry out this test: 
see section 6.1 for a discussion of current data and its limitations. 

\end{enumerate}

In isolated cases,
any of the above observable properties could potentially be explained via
standard astrophysical effects.  But under modified gravity we would expect
to see multiple effects in conjunction, and to see offsets and warps
aligned with the direction of the external gravitational force.
More importantly, we expect these effects to be more prevalent
in unscreened galaxies than in screened galaxies: such a correlation
would be difficult to explain using conventional explanations,
such as bars, spiral arms, asymmetric drift, and external tidal effects.
In the following sections, therefore, we explore aggregate statistics
relating to the presence of these observable effects in screened and
unscreened populations.  

\begin{figure*}
\centering
\includegraphics[width=0.8\textwidth]{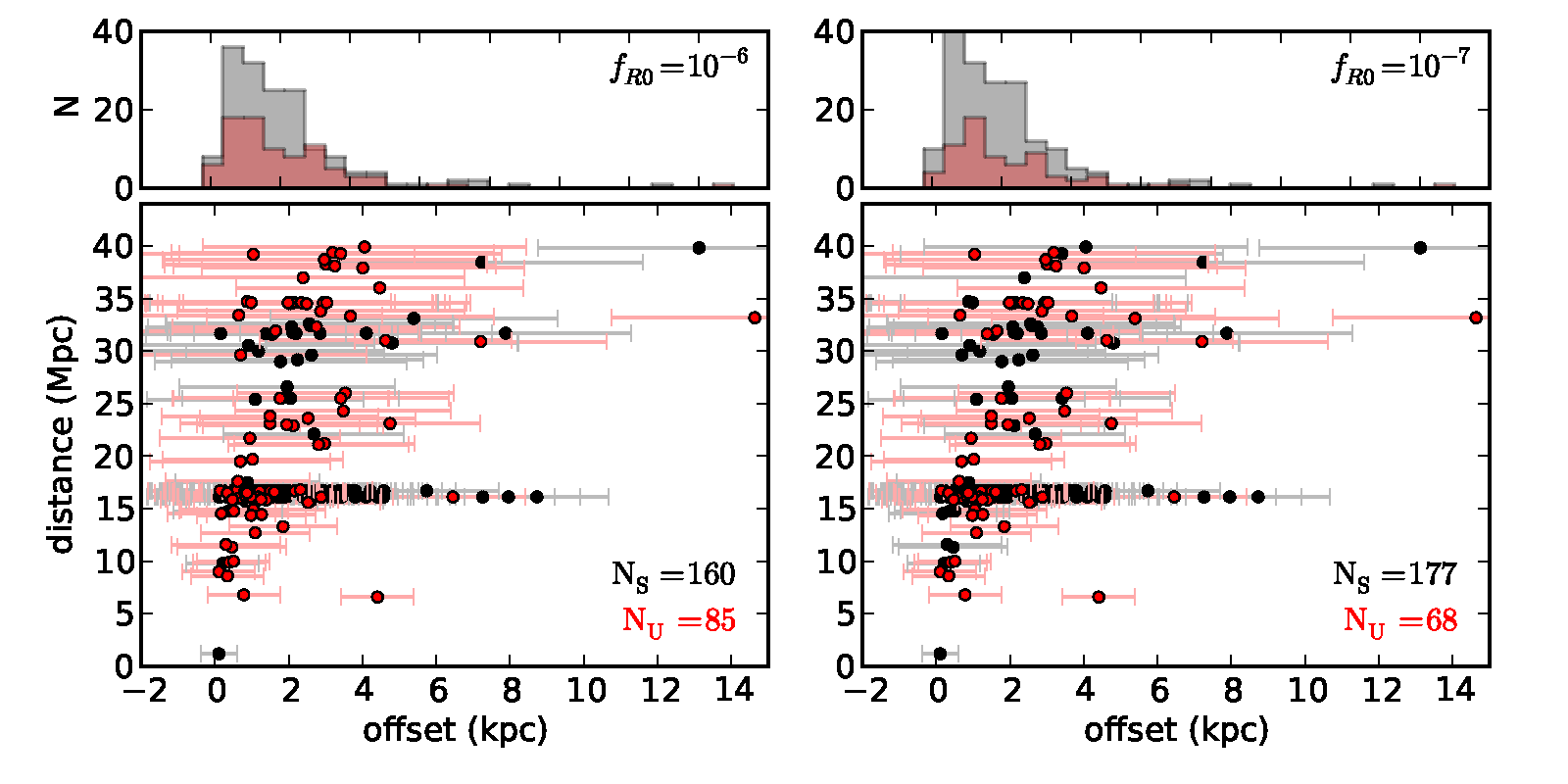}
\caption{The distribution of the measured offset between the HI and
optical centroids for galaxies observed by the ALFALFA and SDSS surveys.
The black (red) points represent screened (unscreened) galaxies for model
parameter values 
$f_{R0}=1\times10^{-6}$ (left panel) and $f_{R0}=2\times10^{-7}$ (right panel)}
\label{fig:offset}
\end{figure*}

\subsection{Observational Tracers}
\label{sec:halpha}

In order to trace gaseous components of dwarf galaxies,
we use observations of the 21 cm line which results from the hyperfine
transition of neutral hydrogen.
In order to trace the stellar component, we use the distribution of optical
light.  Traditionally, \ha{} is also be used as a kinematical tracer
of the disk, but this presents some complications for studies of
modified gravity.

\ha{} emission in dwarf galaxies arises because
the high energy radiation from stars ionizes the neutral gas in the ISM,
forming a  ``Stromgren Sphere''.
As the ions within this region recombine,
the photons emitted in the $n=3\;\to\;n=2$ transition
of Hydrogen lead to the \ha{} emission line.

Historically, this line has been
used to measure the stellar rotation curve, and is expected to give accurate
measurements of the dynamics of the stellar disk in the absence of modified
gravity. However, the possibility of a fifth force which
affects stars and gas calls this expectation into question.
Recall that the star is screened by its own potential, but this potential
may not be strong enough to screen the whole of the surrounding sphere.

A rough calculation shows that for a star of mass $10 M_\odot$ and radius
$5 R_\odot$, the surface potential is $\sim 4\times10^{-6}$. Therefore, this
star will be screened for background field value less than $4\times10^{-6}$.
For $f_{R0} = 1\times10^{-6}\,(2\times10^{-7})$, this potential is only
sufficient to screen ionized gas within
$4.6\times10^{-7}\,(2.3\times10^{-6})$ parsec of the star -- that is, about
4 (20) times the star's radius.
For typical HII regions which have radii of order 10pc, the region screened
by the stellar potential is negligible.  Therefore, we would expect  that the
HII region feels the fifth force and 
orbits faster than the associated star --
this raises interesting questions about what fraction of stars remain
associated with significant HII regions. Thus, under the assumptions of
modified gravity,  it is not safe to assume that \ha{}
traces the dynamics of the stellar component.
We will simply assume that \ha{} is a gaseous tracer and leave 
a detailed study for later work, though we provide a brief empirical
test Section \ref{sec:hihalpha}.

A better tracer of the rotation of the stellar component under modified gravity
may be molecular absorption lines in the stellar spectra, which arise much
closer to the star. Absorption lines such as MgIb or CaII would trace the
stellar rotation curve accurately under both modified gravity and general
relativity. However, these lines are difficult to observe, particularly
in the low-surface-brightness outskirts of the galaxy.

\section{Offset between stellar and gaseous components}
\label{sec:offset}

In this section we consider three tests based on the first observable effect:
the expected offset between unscreened and self-screened components of the
unscreened galaxy sample. Simple calculations indicate that under
reasonable approximations, this offset should be large enough to be
observable \citep{bhuvjake2011}.  We consider three potential avenues toward
this observation below.

\subsection{Offset between HI and optical centroids}
\label{sec:h1-opt-offset}

We first  compare the optical centroid, which traces
the stellar component, to the centroid of HI gas measured via its 21cm emission. 
Here the optical observations are taken from the SDSS $r$-band \citep{aba09}
while the HI centroids
are taken from the ALFALFA survey which used the Arecibo radio telescope
\citep{giovanelli2005, gio07, sai08,ken08,mar09}.
The astrometric precision (about 24 arcseconds) 
of these observations are limited by the
relatively large beam size of the radio maps.
Though the low resolution of the HI image makes any offset difficult to
measure for a single image,
the statistical uncertainty can be reduced by considering a large sample of
galaxies.

To  reduce the  uncertainty in the spatial centroiding,
we limit our sample to those galaxies within 40 Mpc.
To minimize the error introduced by the incompleteness in our catalog described
in Section \ref{pot-sec}, we further  restrict our analysis to the SDSS region,
where  we have information on galaxy groups down to a mass of
$5\times 10^{11}M_\odot$.  Finally, to limit the
effects of different detection thresholds between the SDSS and Arecibo
images, we keep only the galaxies whose optical and HI centroids are
within 1 arcmin.  After these cuts are made, the resulting catalog
contains 245 dwarf galaxies.

In order to identify galaxies which are large enough to be self-screened,
we use the maximum velocity estimated from W50, i.e.\ the
velocity width of the source
line profile measured at the 50\% level of each of the two peaks. This was
corrected for instrumental broadening by the ALFALFA team. However, this must
be further corrected for turbulence and inclination.
The turbulence correction was performed according to \citet{tully85} using an
average velocity dispersion in the gas component of 8 km/s \citep{begum06,
geha06}. Finally, we apply the inclination correction based on the observed
axis ratio $q$ in the SDSS r-band. The maximum velocity is found via:
\begin{equation}
  v_{\rm max} = \frac{W50_c}{2 \sqrt{1 - q^2}},
\end{equation}
where $W50_c$ is the turbulence corrected W50 and $q$ is the axis ratio of
the galaxy. For a background field value $f_{R0}$, the unscreened galaxies
satisfy  \citep[see][]{jainvinu2012}
\begin{equation}
  \left(\frac{v_{\mathrm{max}}}{100}\right)^2 \lesssim
  \frac{f_{R0}}{2\times10^{-7}}.
\label{eq:unscreened}
\end{equation}
The above criteria yields 68 (85) unscreened and 177 (160) screened galaxies
for $f_{R0} = 2 \times10^{-7}$ ($f_{R0} = 10^{-6}$).

In Figure \ref{fig:offset} we show the measured HI/optical
offset for samples of different screening level, along with the
distributions of offsets for two values of $f_{R0}$.
The errors shown are estimated from the 24 arcsec centroid 
error of the radio maps, converted to a physical distance using the
distance to each galaxy.

The interpretation of any observed offset
is made difficult by the fact that there are several potential
astrophysical (non-MG) sources of separation, as well as the fact that we
cannot predict the magnitude or direction of the MG-induced separation.
We address this by proposing an aggregate statistical model of the data,
which takes into account the expected difference between the screened and
unscreened samples.

To account for non-MG related separation, we assume that standard
astrophysical effects will produce a
scatter in the distribution of projected HI-optical separations centered
at zero and with a standard deviation $\sigma_{GR}$.  This scatter will
be present in both the screened and unscreened samples.
To account for MG-related separation, we
assume that within the unscreened sample, there will be an additional
Gaussian scatter centered at zero with standard deviation $\sigma_{MG}$.
Then the observed separation for a screened galaxy and an unscreened galaxy
is respectively
\begin{equation}
  \hat{s}^{scr}_i = abs(s^{scr}_i) + \varepsilon_i \nonumber
\end{equation}
\begin{equation}
  \hat{s}^{unscr}_i = abs(s^{unscr}_i) + \varepsilon_i
\end{equation}
where $\hat{s}^{scr}_i$ and $\hat{s}^{unscr}_i$ are the observed offsets
for the screened and unscreened samples, and
 $s^{scr}_i$, $s^{unscr}_i$, and $\varepsilon_i$ are (unobserved) values
drawn from Gaussian distributions:
\begin{equation}
  s^{scr}_i \sim \mathcal{N}(0, \sigma_{GR}^2) \nonumber
\end{equation}
\begin{equation}
  s^{unscr}_i \sim \mathcal{N}(0, \sigma_{GR}^2 + \sigma_{MG}^2) \nonumber
\end{equation}
\begin{equation}
  \epsilon_i \sim \mathcal{N}(0, \sigma_i^2)
\end{equation}
where  $\mathcal{N}$ denotes a Gaussian distribution and $\sigma_i$ is the
(known)
measurement error associated with observation $i$.

\begin{figure*}
\centering
\includegraphics[scale=0.7]{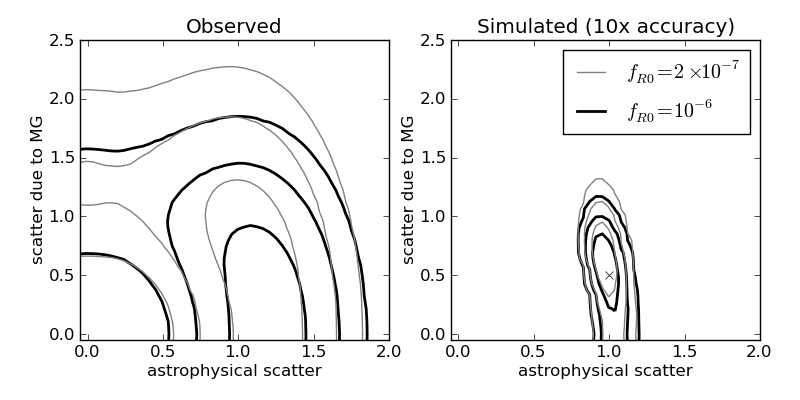}
\caption{The $1$-, $2$-, and $3$-$\sigma$ contours for the scatter in observed
  HI/optical centroid separations in nearby dwarf galaxies.  The left panel
  shows the results given the observed sample of 245 dwarf galaxies described
  in \S\ref{sec:h1-opt-offset}.  Because of the large  beam size of
  the radio maps, the data cannot distinguish between modified gravity and
  general relativity.  The right panel shows a simulated sample drawn from the
  model with a scatter due to modified gravity of $0.5$ kpc (indicated by the
'X'), 
  with a factor of ten improvement in astrometric precision (or a combination of improved astrometry and larger sample size).  The resulting
  posterior can distinguish the predicted effects of modified gravity at the
  1-$\sigma$ level.}
\label{fig:offset-analysis}
\end{figure*}

With this model in place, we can find the most likely parameter combination
within a Bayesian formalism, by writing
\begin{equation}
\begin{split}
  \label{eq:separation_likelihood}
  p(\sigma_{MG}, \sigma_{GR}|\{\hat{s}_i\})
  \propto p(\sigma_{MG})p(\sigma_{GR})
  \left[ \prod_{i \in S^{scr}} g(\hat{s}_i;0, \sigma_{GR}^2 +
\sigma_i^2)\right]\\
\times
  \left[ \prod_{j \in S^{unscr}} g(\hat{s}_j;0, \sigma_{MG}^2 +
\sigma_{GR}^2 + \sigma_j^2)\right]
\end{split}
\end{equation}
where $g(x;\mu,\sigma^2)$ is the Gaussian probability distribution with
mean $\mu$ and width $\sigma$; $p(\sigma_{MG})$ and $p(\sigma_{GR})$ are the
priors on the parameters; and $S^{scr}$ and $S^{unscr}$ are respectively the
sets of screened and unscreened observations.

In Figure \ref{fig:offset-analysis} we show the $1-$, $2-$, and $3-\sigma$
contours
of the posterior distribution given in Eq.~\ref{eq:separation_likelihood}
as a function of $\sigma_{MG}$ and $\sigma_{GR}$.  The left panel shows the
result for the observed data: this data is
not precise enough to rule-out a $\sim 1$kpc scatter due to modified
gravity: this is primarily due to the large observational error in the
measurement of the radio centroid.  In the right panel, we see that
if the aggregate astrometric accuracy were to be improved by a factor of 10
(i.e.\ a 2.4 arcsecond centroid accuracy, 
or a factor of 100 increase in the sample size, or some combination of these
two improvements),
the observations would be sensitive enough to detect a scatter due to
modified gravity of $~\sim 0.5$kpc, on the order of the offsets
predicted by \citep{bhuvjake2011} for typical MG theories.

One important detail of these results for such low signal-to-noise data
is their dependence on
the precise determination of the measurement uncertainty.  Within the model
given by Equation \ref{eq:separation_likelihood}, the measurement error is
degenerate with the astrophysical offset: any increase in assumed measurement
error will be offset by a commensurate decrease in the astrophysical scatter
at maximum likelihood.  Furthermore, because the uncertainty of the measurement
dominates the signal by a factor of up to 4 for each galaxy, even a 10\%
overestimate or underestimate in the centroid error will lead to significant
differences in the maximum likelihood fit.  This caveat should be kept in
mind when interpreting Figure \ref{fig:offset-analysis}.

Thus, while the current data are not precise enough to either confirm or
rule-out modified gravity theories, a large, combined radio/optical survey of
nearby dwarf galaxies with very precise and well-characterized astrometry
could provide strong constraints on typical modified gravity theories.

\subsection{Offset between optical center and \ha{} kinematical center}
\label{sec:kinematical}

\begin{figure*}
\begin{center}
  \includegraphics[width=0.8\textwidth]{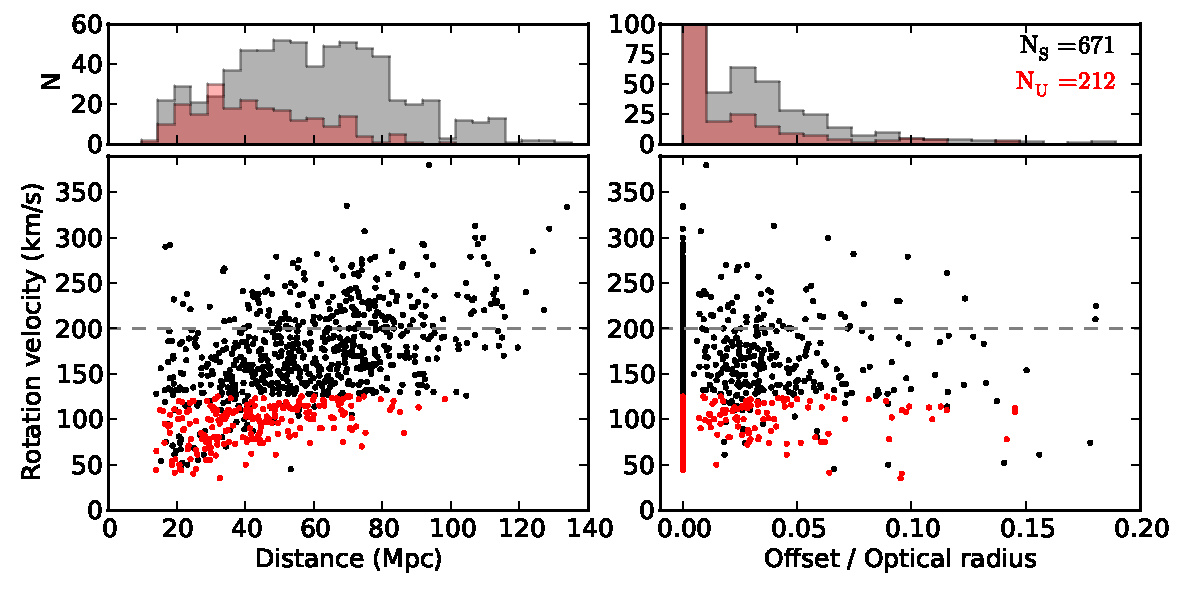}
  \caption{The distribution of galaxies used for the optical/kinematic offset
  test described in Sec
    \ref{sec:kinematical}.  The left panel shows the distance and maximal
    rotation velocity (a tracer of mass) of the galaxy sample; the right
    panel shows the relative offset between the gaseous and stellar components.
    Black (red) points represent screened (unscreened) galaxies at
    $f_{R0} = 10^{-6.5}$.  The clear mass dependence of the screening condition
    is due to self-screening of massive galaxies.}
  \label{offset-kin-den-1}
\end{center}
\end{figure*}

A primary weakness of the measurements explored in the previous section is
the astrometric precision of centroid measurements in the radio.  In
this section we measure the same effect, but using the kinematical
center of the \ha{} disk as a tracer of its centroid.  Here we assume that
the gas producing the \ha{} line is unscreened and traces the dark matter,
which means that the rotation curve should (in principle) have the same
center as the dark matter halo (see Sec \ref{sec:halpha}).
As above, we search for systematic offsets
between the (self-screened) stellar component and this (unscreened) gaseous
component.

For this section, we use data presented by \citet{persic1995}:
they re-analyze 967 \ha{} velocity rotation curves originally 
observed by \citet{mat92}, and publish the offset between the kinematical
center (which symmetrize the rotation curve) and the optical center.  We divide these galaxies into a screened
and unscreened sample using the methodology presented in Section \ref{pot-sec},
leading to 25 (690) unscreened galaxies and 858 (193) screened galaxies
for $f_{R0} = 2 \times 10^{-7}$ ($f_{R0} = 10^{-6}$).

In the left panel of Figure \ref{offset-kin-den-1}, we show the distance to each
galaxy along with its mass indicator, the maximum rotation velocity. The
mild correlation between these measurements can be traced to to observational
bias: smaller, fainter
galaxies are less likely to be observed at larger distances.
Note that velocities range from 50 to 300 km/s, which allow us to test
theories from   $f_{R0}=10^{-6}$ (galaxies self-screened if
$v_\mathrm{max} > 225$ km/s) to $f_{R0}=10^{-7}$ ($v_\mathrm{max} > 70$ km/s).
The color of each point indicates whether the galaxy is screened at a level of
$f_{R0} = 10^{-6.5}$.  There is a relatively sharp boundary between the
screened and unscreened samples seen at $\sim$ 120 km/s rotation velocity: this
is due to the self-screening criterion met by larger-mass galaxies.

Because the size of the expected MG displacement scales with the size of the
dark matter halo \citep{bhuvjake2011}, we work here with the relative offset
$\delta = s_{obs} / r_{opt}$ where $s_{obs}$ is the offset between the optical
and kinematical centers and $r_{opt}$ is the optical radius of the
galaxy.  This relative offset does not show an obvious correlation with
galaxy rotation velocity, as can be seen in the right panel of Figure
\ref{offset-kin-den-1}.

\begin{figure*}
\begin{center}
  \includegraphics[width=0.8\textwidth]{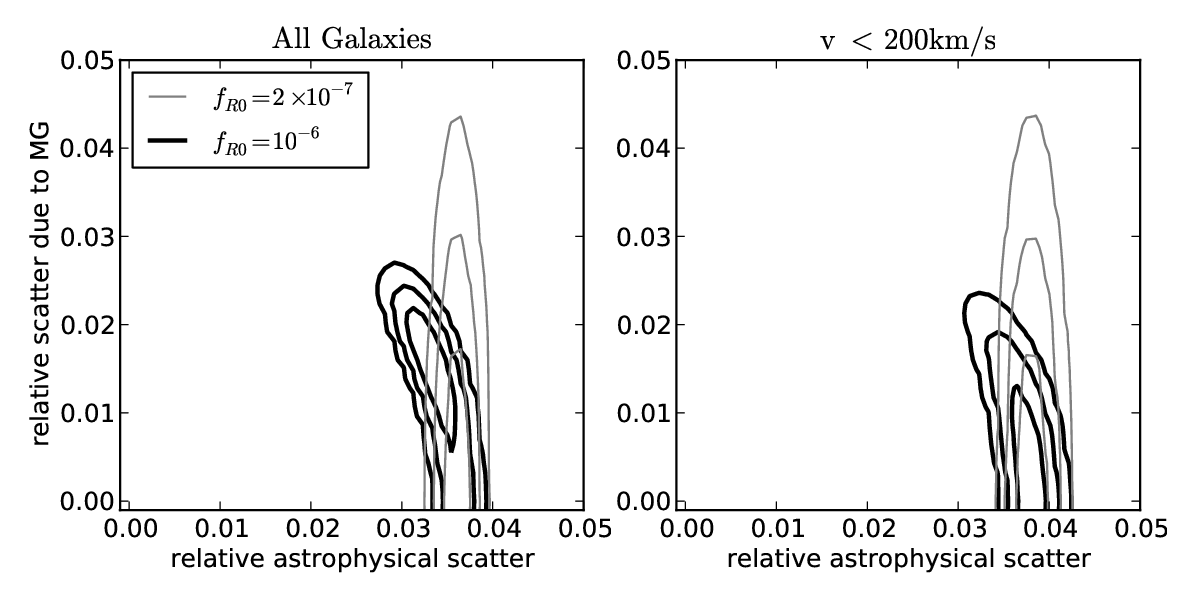}
  \caption{The Bayesian estimate of the contributions of MG and astrophysical
effects to the offset between kinematical and optical centroids
(see discussion of this model in Sec \ref{sec:h1-opt-offset}).  The left
    panel shows a mildly significant offset due to modified gravity:
    we argue that this is due to the bias of the screened sample toward
    high masses.  After controlling for this by removing galaxies
    with rotation greater than 200 km/s, the estimate is
    consistent with GR. 
    }
  \label{offset-kin-den-2}
\end{center}
\end{figure*}

To search for indications of modified gravity, we use the aggregate likelihood
formalism developed in the previous section, replacing the
absolute deviations $s$ with the relative deviations $\delta$.  The resulting
posterior is shown in Figure \ref{offset-kin-den-2}.  In comparison to the
ALFALFA results in Figure \ref{fig:offset}, the data posterior is
much tighter: this is mainly due to the increased astrometric accuracy
of these measurements.

The left panel of Figure \ref{offset-kin-den-2} shows results for the entire sample, and
(for $f_{R0} = 10^{-6}$) indicates an approximately $1-\sigma$ detection
of an offset due to modified gravity.  Because this result uses the whole
sample, however, it is potentially biased by the sizes of the galaxies in
the sample. As shown in Figure \ref{offset-kin-den-1},
the vast majority of screened galaxies
in our sample are not environmentally screened, but self-screened.  The
differential offset shown in this analysis, then, not only reflects the
difference between screened and unscreened galaxies, but also
the difference between large and small galaxies.  Even absent modified
gravity effects, we may expect statistically significant differences in
observed offsets between these samples.

To limit the effect of this size bias, we repeat the analysis, limiting our
galaxy sample to those with a maximum rotation velocity less than
200 km/s (i.e.\ those galaxies below the dashed line in
Figure \ref{offset-kin-den-1}.  Having controlled for this bias, we see in
the left panel of Figure \ref{offset-kin-den-2} that the offsets are consistent
with General Relativity for both screening levels.

Still, this analysis does not rule out offsets due to modified gravity
with a scatter of approximately 2\% of the optical radius of the galaxy.
Under ideal conditions, \citet{bhuvjake2011} showed that the expected
offsets for typical modified gravity theories could be as high as 10\% the
optical radius of the galaxy.  This predicted offset decreases as the size
of the galaxy increases, and as the distance to the neighboring galaxy
increases.  Thus a 2\% upper-limit is not enough to definitively
rule-out modified gravity theories.

\subsection{Offset between centroids of red giants and main sequence stars}
\label{sec:rgb}
In modified gravity scenarios, red giant stars present an interesting case:
their cores may be dense enough to self-screen, while their diffuse outer
shells remain unscreened and feel the fifth force.  This partial screening
can lead to interesting effects: for example, 
\citet{changhui} have recently shown that red giant stars evolve differently
under modified gravity compared to main sequence stars, and that this
difference can be used to constrain modified gravity theories.

In this section we make use of red giant populations
to constrain modified gravity in a different way:
just as unscreened gas and dark matter will separate from
the screened main sequence stars in the presence
of an external potential gradient, partially screened red giants
will be expected to separate from the population
of main sequence stars as well.

\subsubsection{Data and centroid estimation of RGB and main sequence stars}
Recently, \citet{dalcanton09} published positions and magnitudes
of individual stars within 69 nearby dwarf galaxies
(distance $<$ 4 Mpc , $|b| > 20$ deg)\footnote{See
\href{http://www.nearbygalaxies.org}{http://www.nearbygalaxies.org}}.
Absolute magnitudes of these galaxies ($M_B$) range from -20 to -8. They  
observed nearly $10^3$ to $10^5$ stars in each galaxy.
In our study we need full images of the galaxies to estimate the displacement
between RGB (red giant branch) and main sequence stellar populations as partial
images of galaxies introduce artificial displacement between those two
populations. We use only those galaxies which pass this criterion. In
addition, we exclude all the galaxies in which foreground contamination
leads to poorly-defined red giant branches.
Finally, we include galaxies only if they have at least 3000 observed stars.
The final sample consists of 28 galaxies: under either screening assumption
($f_{R0}=10^{-6}$ and $f_{R0}=10^{-7}$) every galaxy is unscreened,
which means this test cannot rely on a screened control sample.

For each galaxy, we select red giants from the 
color-magnitude diagram. In Figure \ref{fig:selectrg} we show an example of
red giant star selection.
Because the fifth force acts only on the outer regions of the largest giant
stars, we tune the cutoff to select only the brightest, and therefore the
largest, of the red giant stars.
Next we define the area of the galaxy within which
the centroid will be calculated.
This is done by taking isocontours of stellar number density significantly
above the noise level.
Errors in the centroids of the red giant and main sequence populations
are calculated using bootstrap resampling with 50 realizations.
These are shot-noise dominated.
Since the number of red giants is at least an order of magnitude
lower than the number of main sequence stars, the error in the centroid
displacement is dominated by the error in the centroid position of
the red giant sample.

\subsubsection{Estimation of the external fifth force}
Because of the lack of control sample, detecting modified gravity here
depends on correlating the direction of displacement with the strength
and direction of the background gravitational potential gradient at
each dwarf galaxy due to its neighbors.  In order to estimate this,
we use the
catalog of  nearby galaxies given by \citet{kar04}. This catalog is nearly
complete out to a distance of 10 Mpc and it contains a total of 431 galaxies
with known mass for 313 galaxies. We fit a linear relation between the
mass and magnitude (in logarithmic scale) to the galaxies with known mass and
use this empirical relation to estimate the masses of the remaining galaxies.
By using distances and masses of neighboring galaxies we can construct the
force vector (per unit mass) as 

\begin{equation}
\vec{a} =\sum_{i \in {\rm unscr}}{\frac{G M_i}{|\vec{r_i}
-\vec{r}|^2} \frac{\vec{r_i}
-\vec{r}}{|\vec{r_i} -\vec{r}|} }
\label{eq:accelaration}
\end{equation}
where $m$ is the studied galaxy mass at position $\vec{r}$,
 $M_i$ the mass of the
neighbor $i$, which is at position $\vec{r_i}$ and G is Newton's
constant. The sum includes only unscreened neighbors at a distance below the
Compton length. 
The gravitational potential W can be calculated as 
\begin{equation}
 W=\sum_i{\frac{G M_i}{|\vec{r_i} -\vec{r}|}  }
\end{equation}
where the sum includes all neighbors at a distance below the Compton length.
Here we use the potential to estimate the magnitude and direction of the
background gravitational force, as well as to estimate the potential to
define the screening level for the dwarf galaxy sample.
The force has some errors due to the
uncertainty in distance (5\%) and mass ($\sigma_{log(M)}=0.2$).
In our case, uncertainty in mass is the
dominant source of error. 

\begin{figure*}
\includegraphics[scale=0.5]{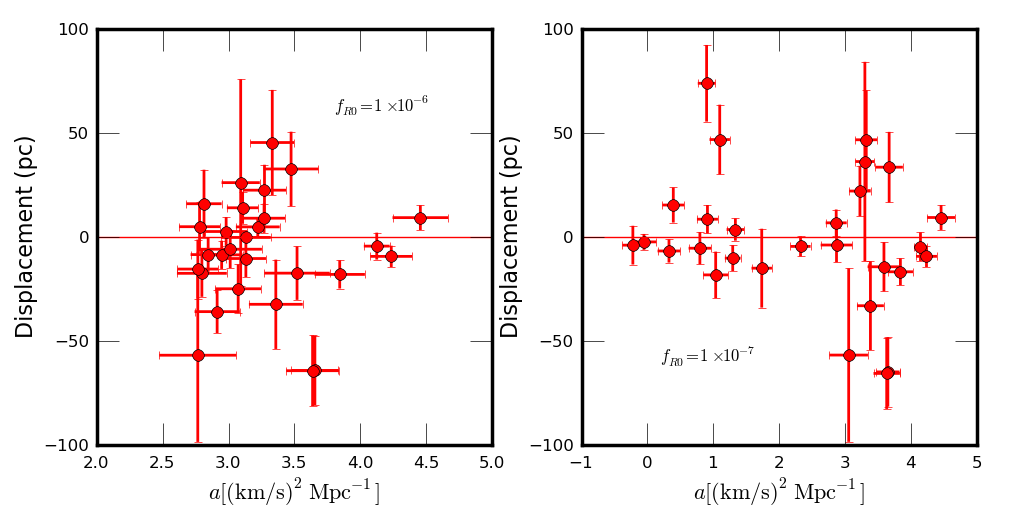}
  \caption{The displacement of red giants from main sequence stars projected along 
the external fifth force from a sample of 28 nearby dwarf galaxies. The displacement is plotted as a function of the external force per unit mass
(Eq. \ref{eq:accelaration}). 
In the left (right) panel, we set $f_{R0}=10^{-6} (10^{-7})$. 
All the galaxies are unscreened. There is no significant correlation between
the
displacement and the applied force.
 The mean displacement in the direction of the 
external force is $-10.2\pm5.5$ pc  and $-4.7\pm6.4$ pc for
the two cases, consistent with GR. 
\label{fig:rgdispprojforce}}
\end{figure*}

\subsubsection{Results and comparison with theory}
In Figure \ref{fig:rgdispprojforce} we plot the displacement of red
giants from  main sequence stars projected to the external force vector as a 
function of the force per unit mass. The force vector is previously
projected to the plane of the sky perpendicular to the line of sight.
In order to calculate the error in the projected displacement, 
we take into account both the error in the centroids and the error in the
vector force, and show the results for $f_{R0}=10^{-6}$ and $f_{R0}=10^{-7}$.
There is no statistically significant correlation between the forc and
projected displacement.
The weighted average displacement in the direction of the 
external force for $f_{R0} = 10^{-6}$ ($f_{R0} = 10^{-7}$) is
$-10.2\pm5.5$ pc ($-4.7\pm6.4$ pc),
consistent with zero to within 2-$\sigma$ (1-$\sigma$).
The errors on these estimates include intrinsic dispersion, as explained
in Appendix \ref{ap:avg}.



In Figure \ref{fig:rgdisppred} we compare the projected displacement to the
prediction, given the external force and assuming a cored isothermal sphere
(see Equation 2.3 of \cite{bhuvjake2011}
for the analytical expression of the offset).
The prediction takes into account errors in the masses and distances of
neighboring galaxies, as well as errors in the determination of the
galaxy mass, core radius, and core mass. The last two quantities
are estimated using the strong empirical correlation
between virial mass and core radius and core mass \citep{swaters03,swaters11}.
We have assumed $\Delta G/G =1$
for the predictions, but this result can be easily scaled for other values of 
$\Delta G/G$.
The mean difference between the observed displacement $d$ and predicted
displacement $p$ is $p-d=12.5\pm6.1$ pc ($-0.7\pm4.8$ pc) for
$f_{R0} = 10^{-6}$ ($10^{-7}$).

We find that both these tests are unable to rule-out either GR or modified
gravity. The main reason for this lack of
discriminatory power is because most of the galaxies are very isolated,
so the external force has a small magnitude.  In Appendix 
\ref{ap:rgb-systematics} we discuss in detaila series of important
systematic uncertainties involved in this test.


\begin{figure*}
\includegraphics[scale=0.5]{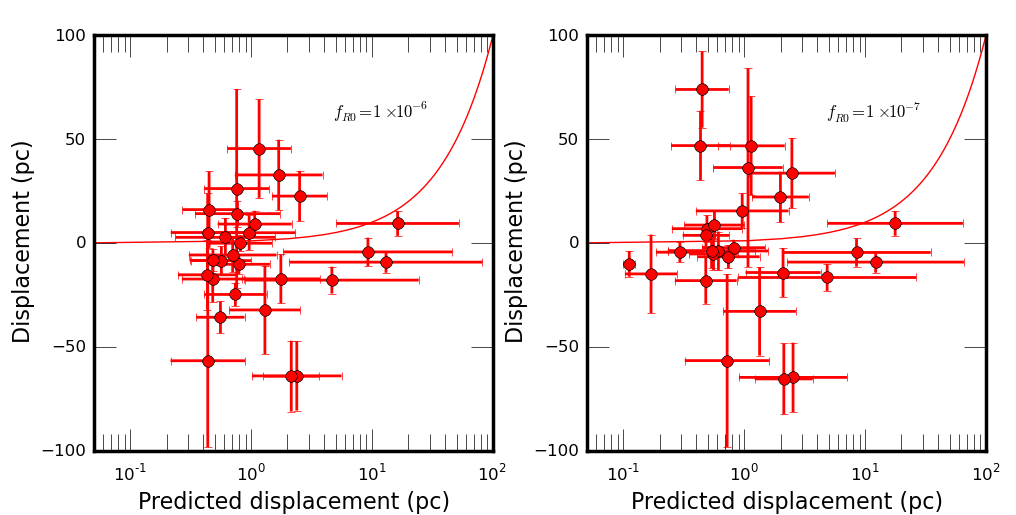}
  \caption{The estimated projected displacement, as in Fig. 5, is compared with the predicted displacement assuming $\Delta
G/G=1$. In the left (right) panel, we set $f_{R0}=10^{-6} (10^{-7})$. 
 The solid
line shows the one to one correspondence. It is evident that while 
the observations are consistent with
GR, one cannot rule out modified gravity 
given the large errors in the predictions and systematics.}
\label{fig:rgdisppred}
\end{figure*}

\subsubsection{Future directions}
We find that tests of modified gravity through
comparison of the positions of red giant and main sequence stars
are inconclusive, mainly owing to the error associated with the
small sample size.  Also, because red giants are only
partially screened, the signal may be diluted when compared to
the tests of the previous sections.
Additionally, the lack of a control sample means that
the uncertainty in the external force is the limiting factor
if one wants to constrain modified gravity with only the unscreened sample as
such tests require much better prediction to compare the data. 

Future studies making use of this effect should focus on unscreened galaxies
with very close neighbors, which will thus experience stronger external
potential gradients, and drive the separation to more detectable levels.
It would also be helpful to focus on similarly-sized screened galaxies to
provide a control sample (though this pursuit may be limited by the relative
paucity of nearby screened dwarf galaxies).  
The best approach would be to focus on a limited number of galaxies with
regular morphology, to reduce intrinsic systematic effects, with a very well defined
external force.

Another potentially interesting method 
for this sort of measurement would be to use integral field
spectra of galaxies and estimate the red giant and main sequence contributions
to the spectrum as a function of position: any systematic offset, aligned with
the external fifth force in the inferred mass of red giant and main sequence
stars could indicate the presence of a fifth force acting on the red giant
population.

\section{Warping of edge-on galaxies}
\label{sec:warp}

In the previous section, we discussed the expected offset of the stellar
and dark matter components of suitable unscreened dwarf galaxies within
modified gravity.  As shown in \citet{bhuvjake2011}, this offset may lead
to an observable morphological signature: a bowl-shaped warp in the stellar
disk, aligned with the direction of the external gravitational force and this
effect will be maximized when the gravitational
force is perpendicular to the plane of the disk. When viewed edge-on,
this will result in a symmetric U-shaped warp in the disk of the infalling
galaxy.  Because standard astrophysical dynamics (e.g. tides) could lead to
similar effects, we use screened galaxies as a control sample and look for
a statistically significant higher fraction of such U-shaped warps in the
unscreened sample.

\subsection{Data and estimation of warp parameter}
We search for warps in low mass, nearly edge-on galaxies using the
ALFALFA and SDSS data. The rotation velocity measured by the former is used to
find self-screened galaxies, and images from the latter are
used to estimate the galaxy warp.
Because warp is difficult to observe in face-on
galaxies, we limit our sample to those galaxies with an axis ratio
of less than 0.6. In the resulting sample of 495 galaxies, 128 (59) 
galaxies are screened and 367 (158) unscreened, for a Compton scale length of 3
(1) Mpc which corresponds to $f_{R0}=1 \times 10^{-6}$ ($f_{R0}=2 \times
10^{-7}$). For $f_{R0}=2 \times 10^{-7}$ we use only galaxies with
$v_\mathrm{max} < 200$ km/s in the screened sample.

\begin{figure*}
\begin{center}
  \includegraphics[scale=0.5]{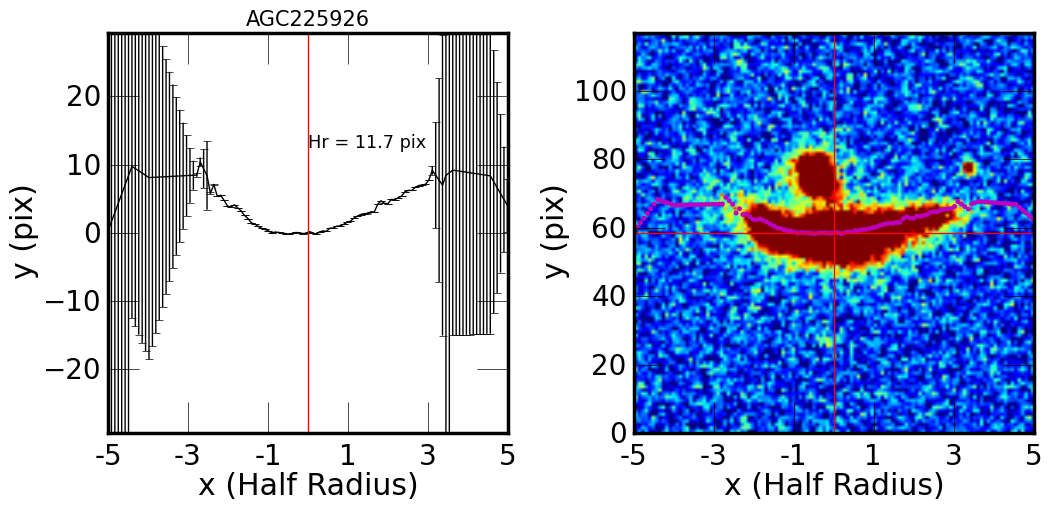}
\caption{An example of the estimation of warping from a disk galaxy image. 
The left panel shows the estimated warp curve. The red
vertical line shows the $x$-center of the galaxy. The $x$-axis is given in the unit
of half light radius ($h_r$) of the galaxy. In the right panel we superimpose
the warp curve on the optical image of the galaxy. 
}
\label{warp-eg}
\end{center}
\end{figure*}

In order to reduce the type of mass-related systematic effects discussed in
the previous  section, we include only smaller galaxies in which the
self-screening condition (Eq.\ \ref{eq:unscreened}) does not apply: the
separation of screened and
unscreened galaxies is based only on the environment.
The warp of the galaxies is found by slightly modifying the method used
in \citet{jim97}: after sky subtraction, we first rotate the SDSS images
so that the major axis of the galaxies align with the horizontal axis.
The center of rotation is found using the SExtractor package \citep{bertin96}.
We mask stars and other contaminating
nearby objects in the image. After removing all the contaminating objects and
pixels in the frame we find the centroids
for each column in the image by fitting a Gaussian to the column and taking the
first moment. This leads to a `warp curve', i.e. the centroid of the
columns as a function of distance from the center of the galaxies.
To clean the curve, we starting from the center of the galaxy and
sweep through the curve, omitting any column which has a linearly-interpolated
centroid that deviates more than three pixels from that of the previous column.
An example of such a cleaned warp curve is shown on Figure \ref{warp-eg}.

After estimating the warp curve, we quantify the overall
warp by defining the following two measures:

\begin{equation}
w_1 =  \frac{\int_0^L \frac{x}{L} \frac{y}{L} \frac{dx}{L}}{\int_0^L
\frac{x}{L}\frac{dx}{L}}=\frac{2}{L^3} \int_0^L x y dx
\label{warp-param} 
\end{equation}

\begin{equation}
w_2 = \frac{\int_0^L{x y dx}}{\int_0^L{x dx}}=\frac{2}{L^2} \int_0^L x y dx
\label{warp-param2} 
\end{equation} 
for the left ($x < 0$) and right ($x > 0$) side of the galaxy.

$w_1$ is dimensionless and proportional to the mean $y/L$ displacement from the
$x$-axis weighted by the distance from the center ($x/L$),
where $L = \max(|x|)$.
The normalizing factor ($1/L$) assures that $w_1$ is independent
of the choice of the measurement unit.
$w_2$ has units of kpc, and measures the average deviation from a
horizontal plane of the galaxy weighted by $x$ (horizontal axis).
In each warp measure, the error is calculated via standard propagation
of the width of each column centroid.

We correct both warp measures
for the inclination of the galaxy based on \citet{geha06}.
This is done by dividing the measured warp curve by $\sin(i)$ where $i =
\sqrt{\frac{1 - (b/a)^2}{1 - 0.19^2}}$ for a relative disk thickness of 0.19.
The warp is estimated within three half-light radii of the galaxy (beyond
this radius, the images tend to become very noisy), though the results of
this section do not vary
appreciably as this limit is changed from 2 to 4 half-light radii.
In Appendix \ref{ap:warp-distri} we define a third warp measure,
and show that the following results are unchanged.

\subsection{Results}
\begin{figure*}
\begin{center}
 
\includegraphics[scale=0.45]{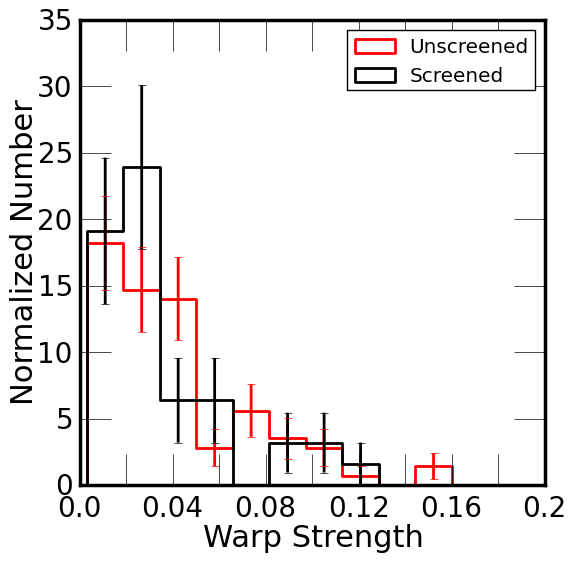}
\includegraphics[scale=0.44]{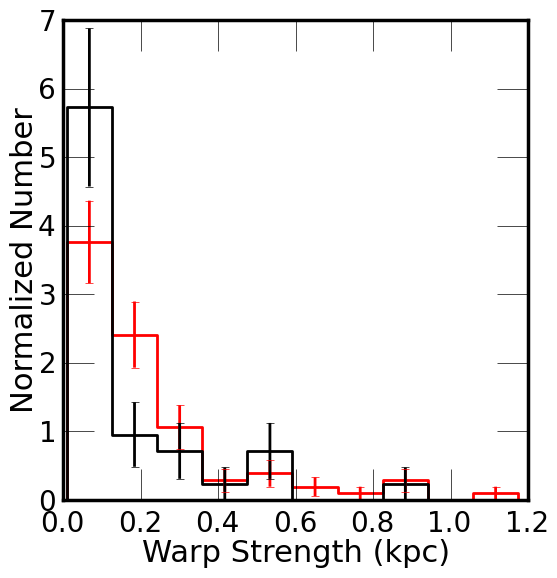}
\caption{ The distribution of warps for screened and unscreened galaxies for
$f_{R0} = 2\times10^{-7}$.
The left panel shows the dimensionless warp estimated using 
Eq.\ref{warp-param} and the  right panel shows the estimate in kpc using   
  Eq.\ref{warp-param2}. No significant difference is found
between the distributions of warps of  screened and unscreened galaxies in 
the left panel. A marginal difference is found on the right with more warp
for unscreened galaxies. This discrepancy does not remain if we account for size
bias as discussed in the text.}
\label{fig:warp-distr}
\end{center}
\end{figure*}

We classify the warp as ``U-shaped'' if $w_1$ or $w_2$ has opposite signs on
the left
and right sides, and ``S-shaped'' if the signs match.
To quantify the warp strength, we average the absolute values of the
warp in the left and right sides of the galaxy.
Figure \ref{fig:warp-distr} shows the normalized distribution of $w_1$ 
(left panel) and $w_2$ (right panel) 
for screened and unscreened galaxies displaying a
U-shaped warp. The errors in the histograms are estimated using Poisson
statistics in each bin.

It is expected within modified gravity that unscreened galaxies with
close neighbors will preferentially show large U-shaped warps.
A K-S test shows that the $w_1$ distributions are similar between the
screened and unscreened samples: this implies that the appearance of
warp does not depend on the screening level.
However, we find that $w_2$ distributions are
marginally different: the unscreened galaxies are slightly more warped. The
null hypotheses that the screened and unscreened distributions are identical is
accepted only with a probability $\sim$ 0.88.
In Appendix \ref{ap:distance-systematic} we show that this difference can be
attributed to size or
distance bias. In Appendix \ref{ap:warp-distri} we show that
warp distributions remain similar even if we expand the sample to include
all types of observed warps.

\begin{figure*}
\begin{center}
\includegraphics[scale=0.43]{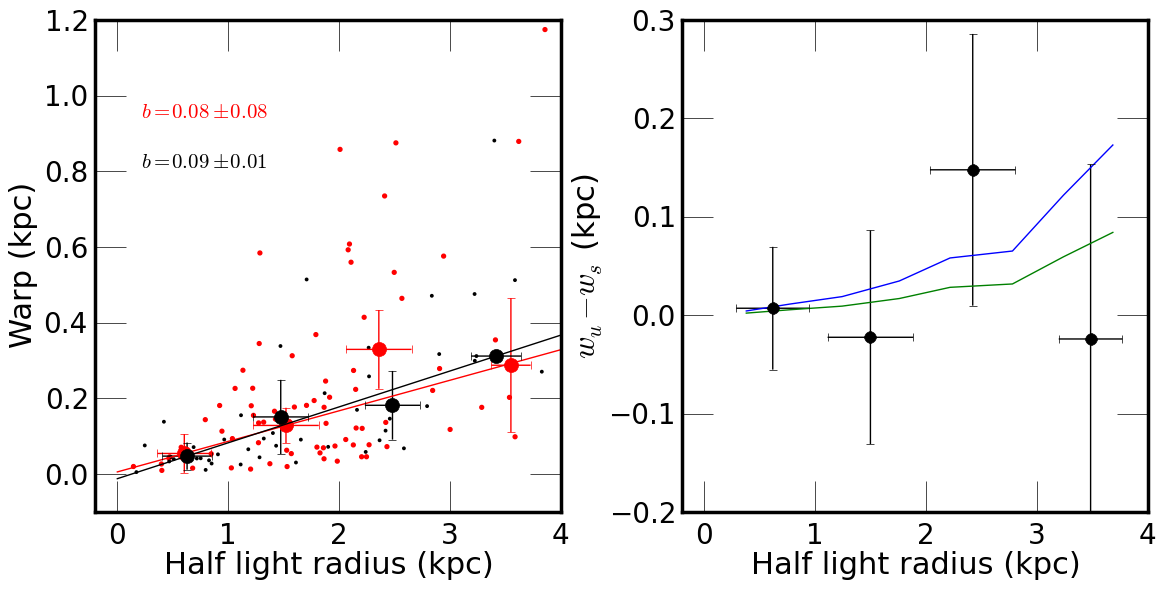}
\caption{The left panel shows the strength of the warp in kpc as a function of
galaxy size. The small black and red dots
represent individual measurements of screened and unscreened galaxies for
$f_{R0} = 2 \times10^{-7}$. The large circles
with error bars show the average warp.  The
lines are the best fit linear relation to these averages. The slopes of
the lines are shown in the top left corner. The right panel shows the estimated
contribution to the warp due to MG, i.e., the difference between warp
of unscreened
($w_u$) and screened samples ($w_s$). The blue and green lines shows the
predictions with $\Delta G/G = 1$ and $\Delta G/G = 1/3$. Given the error
bars, we are unable to test these models. 
}
\label{fig:size-warp-kpc}
\end{center}
\end{figure*}

The distance-dependent effect, described in Appendix
\ref{ap:distance-systematic}, suggests that the ideal way to
compare the $w_2$ of screened and unscreened samples is to consider
only galaxies of similar physical size.
The left panel of Figure \ref{fig:size-warp-kpc} shows the
warp $w_2$ versus the half light radius of galaxies. As expected, the
warp is larger for bigger galaxies. Small points indicate measured
values, and the large points show the average and dispersion within
0.5kpc bins in half-light radius (see Appendix \ref{ap:avg}).
The lines shows the best fit linear relation between the
half light radius and the warp, with slopes listedin the top left corner.
We see that when comparing galaxies of similar size, the warp is
equivalent: this indicates that the difference of distributions in the
right panel of \ref{fig:warp-distr} as due to this size bias. 

\subsection{Comparison with theory}
Our analysis of galaxy warp shows that there exists no significant
difference between the occurrence of warps in screened and unscreened
samples. Here we use this non-detection to place 
limits on strength of the fifth force. Given the low levels of observed warps, 
we assume that any contribution 
due to modified gravity will be added linearly to astrophysical effects
which are estimated from the screened sample: under this assumption, to
determine the warp associated with MG, we subtract the screened value from
the unscreened value.

We find the average warp ($\langle w_1\rangle$) and error on the average 
using the method described in Appendix \ref{ap:avg}. We find that $\langle
w_1\rangle=0.035 \pm 0.014$ for unscreened galaxies and $\langle
w_1 \rangle=0.032 \pm 0.020$ for screened galaxies (i.e. the
warp due to astrophysical process is a 3\% effect) for $f_{R0} =
2\times10^{-7}$. This implies that the estimated warp in a modified gravity
scenario ($\hat{w}_{MG}$) is as low as $0.003 \pm 0.024$ (0.3\%). See Table
\ref{tab:warp} for
the results of other background field values. We can use this in conjunction
with a theoretical expectation of the warp strength to place loose
constraints on modified gravity. This expected warp can be estimated
as described by \citet{bhuvjake2011}; the result 
is sensitive to the mass profile of the galaxy,
as well as the strength of the external potential gradient. 
Unfortunately, both these are difficult to determine from the data.

To estimate the external potential gradient,
we follow \citet{bhuvjake2011} and assume that a galaxy will
acquire a velocity of $\sim 100$ km/s in 3 Gyr.
This gives an average acceleration of $\sim 10^{-15}$ km/s$^2$
We also need to consider
the direction of this force at the galaxy to get the effective force to
generate the warp. We marginalize over the angle by averaging
1000 random angles between -90 deg and 90 deg. The mass at a given
radius of the
galaxy is found by the assumption of cored isothermal sphere. Based on these
assumptions we generate warp curves for these galaxies.
This analysis leads to a theoretical prediction of the modified gravity
warp contribution of
$w_{MG}= 0.003_{-0.001}^{+0.003}\,(w_{MG}= 0.006_{-0.003}^{+0.006})$
for $\Delta G/G = 1/3\, (\Delta G/G = 1)$,
with errors indicating the scatter in predicted values.

The right panel of Figure \ref{fig:size-warp-kpc} shows the average
values of $w_2$, as well as the theoretical predictions outlined
above, as a function of galaxy size (these values are tabulated
in Table \ref{tab:warp}).
The blue and green lines show the expected warp for $\Delta G/G
= 1$ and $1/3$, respectively.
The large error in the measured values
is due primarily to the fact that we are looking for a
relatively small effect in a small sample of galaxies.
The figure indicates that the measured warp is not precise enough to
distinguish between modified gravity and GR, which corresponds to zero
warp in this figure.


\subsection{Future Directions}
Our study of galaxy warps cannot rule-out either the predictions of
MG or of GR.  The two limitations are the relative
scarcity of objects with kinematic data (like ALFALFA), and the uncertainty
in halo parameters that goes into the warp prediction.

Based on our estimates of
uncertainties in the measured and predicted warps we can estimate
the number of galaxies with kinematic data required to
reduce the error in the observed warp by a factor of 8 (so that
the measurement error is comparable to the uncertainty in the predictions). 
We need $\sim 8,000$ dwarf galaxies to test $f_{R0} =
2\times10^{-7}$ and $\sim 20,000$ galaxies for $f_{R0} =
1\times10^{-6}$ (calculated from third and fourth lines of Table
\ref{tab:warp}). This assumes that the halo parameters have the same accuracy as we have in
this paper.  Note also that even though the number of
galaxies required to test $f_{R0} = 2\times10^{-7}$ is a factor of two
less than that of $f_{R0} = 1\times10^{-6}$, those galaxies should be
very low mass and isolated, and therefore will require deep surveys.
Another avenue of improvement is to obtain better optical images from
current surveys (e.g Dark Energy Survey). This imaging should be supplemented
by spectroscopic or 21cm surveys which can measure rotation velocities.

\section{Stellar and gaseous velocity rotation curves}
\label{sec:rotationcurves}

In this section we explore the use of kinematic measures to detect
effects of modified gravity.  There are two classes of expected effects,
both predicted by \citet{bhuvjake2011}.
First, in the presence of a background potential gradient, a galaxy
with a rotation axis perpendicular to the gradient will exhibit a distorted
stellar rotation curve.  Second, within the galaxy, the unscreened material
will feel a stronger force and thus have its rotation curve enhanced by
a factor of $\sqrt{1 + \Delta G/G}$.  Both these measures depend on finding
an accurate tracer of the kinematics of main sequence stars.  We discussed
\ha{} as a potential tracer in Section \ref{sec:halpha}; we return to this
subject here before proceeding with one of these tests.

\subsection{HI-\ha{} velocity difference}
\label{sec:hihalpha}

Historically, \ha{} emission has been assumed to trace the stellar rotation
curve.  Because it originates in HII regions near hot stars, this assumption
is well founded.  However, as we argued in Section \ref{sec:halpha}, this
assumption may not hold in a modified gravity scenario.  Because of the
potentially complicated dynamical scenario of stars which lag behind the HII
regions they create, however, simple arguments such as these may be incomplete.

In this section, we address this question empirically
by comparing the \ha{} rotation curve to that of HI, which is
associated with unscreened gaseous component. Any similarity between
these two within unscreened galaxies would indicate the safety of our
assumption that \ha{} emission traces the unscreened galaxy component.

\begin{figure}
\begin{center}
  \includegraphics[scale=0.4]{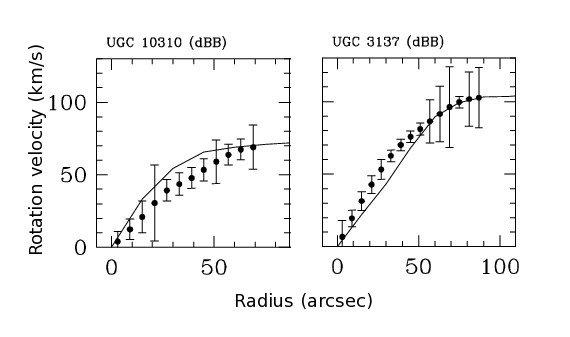}
\caption{An example of rotation curves for two galaxies, measured in \ha{}
(dots with error bars) and HI (solid lines).
The data are taken from \citet{Swaters2009}.}
\label{h1-ha-rot-eg}
\end{center}
\end{figure}

In order to estimate the difference between HI and \ha{} rotation
curves, we extract HI and \ha{} data from Figure 3 of \citet{Swaters2009}, which
contains 22 galaxies. The errors in $v_{HI}$ were obtained by private
communication with Swaters.  Each HI observation has one or two \ha{}
counterparts (see the references therein); in addition, we have \ha{} data for
14 galaxies from GHASP. Note that all galaxies in the sample can safely be
assumed to be unscreened due
to their low mass and lack of massive neighbors. In Figure \ref{h1-ha-rot-eg} we
show the comparison of HI and \ha{} rotation curves for two galaxies taken from
\citet{Swaters2009}: the galaxy in the left panel has HI $>$ \ha{}, while the
galaxy in the right panel has HI $<$ \ha{}.

\begin{figure}
\begin{center}
\includegraphics[trim= 0cm 0cm 0cm 0cm, clip = true,
width=0.4\textwidth]{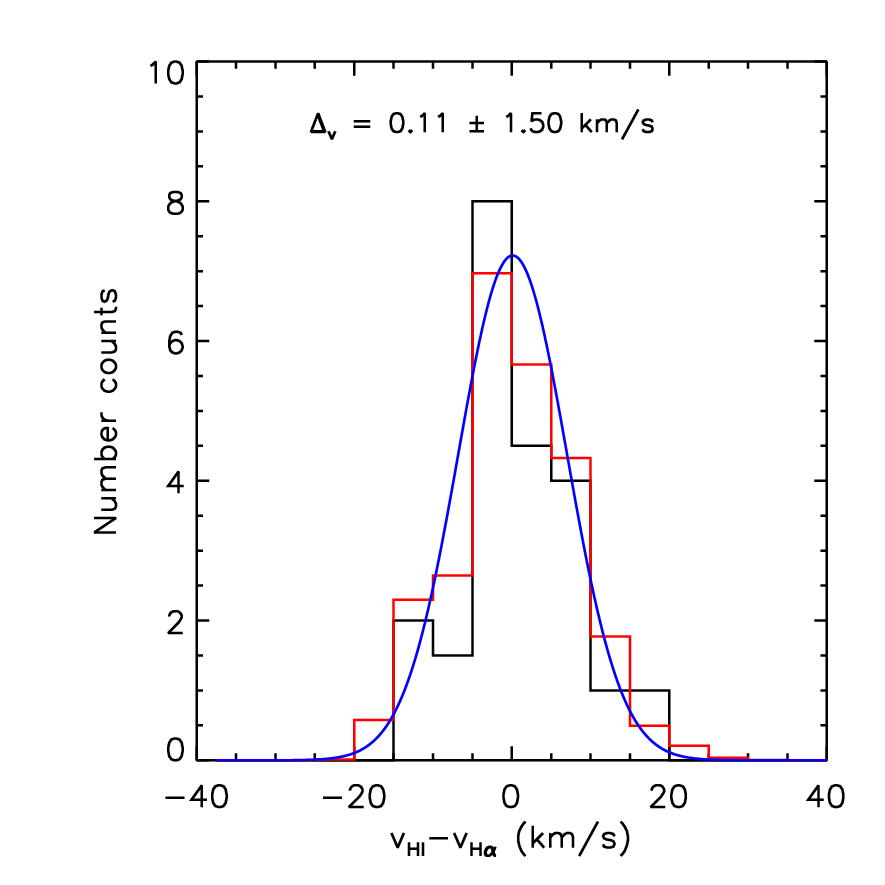}
\caption{The difference between HI and \ha~ rotation velocities.
We plot the distribution of $\Delta {v}=v_{HI}-v_{H\alpha}$ for the 
\citet{Swaters2009} data (black histogram). The red histogram includes 
individual errors in the $\Delta v$ measurements. We over plot in
blue the best fit Gaussian. The mean value of the difference is indicated as well, showing that the 
rotation curves agree
to better than 2 km/s.}
\label{h1-ha-rot}
\end{center}
\end{figure}

To quantify the difference between HI and \ha{} rotation curves statistically
we first average both sides of the HI and \ha{} rotation curves
separately about a common center to reduce the observational uncertainty in the
measured rotation curves, and then compare the HI and \ha{} curves to estimate
the weighted average velocity difference, $\Delta v = v_{HI} - v_{H\alpha}$.
If both \ha{} and HI rotate similarly, then the $\Delta v$ will
be approximately normally distributed about zero.
In Figure \ref{h1-ha-rot} we plot the distribution of $v_{HI}-v_{H\alpha}$ in
black for \citet{Swaters2009} data. When more than one
\ha{} measurement is provided for the same galaxy, we weight equally the
contribution of each  $\Delta v$ to the mean. Accounting for errors in the
observed values gives the red histogram.\footnote{
In order to take into account individual errors and
intrinsic dispersion to calculate the mean and the error we follow the
procedure described in Appendix \ref{ap:avg}.}  The best-fit Gaussian is
indicated by the blue curve.

We obtain $\langle \Delta v\rangle=0.11\pm 1.50 \pm 5$ km/s (statistical and
systematic error) when we use HI and \ha{}
from Figure 3 of \citet{Swaters2009}.  Combining different \ha{} observations
and testing different ranges in the velocity curve leads to a 2 km/s
contribution to the systematic error; the remaining 3 km/s comes from the
potential for asymmetric drift \& non-circular motion to affect the
relationship \citep{Swaters2009, deblok2002}. The errors
additionally include those due to both measurement and non-circular motions,
which we found by taking the difference between approaching and receding
sides. See \citet{Swaters2009} for details. 

With these affects accounted for, we find that the HI and \ha{} rotation
curves are consistent.  This consistency confirms that it is safe to assume
that \ha{} traces the unscreened, gaseous component of the galaxy even
in the presence of a fifth force.

\subsection{Asymmetry in \ha~ rotation curves}\label{sec:asymmetry}

\begin{figure}
\begin{center}
\includegraphics[trim= 3cm 0cm 0cm 0cm, clip = true, width=0.45\textwidth]
{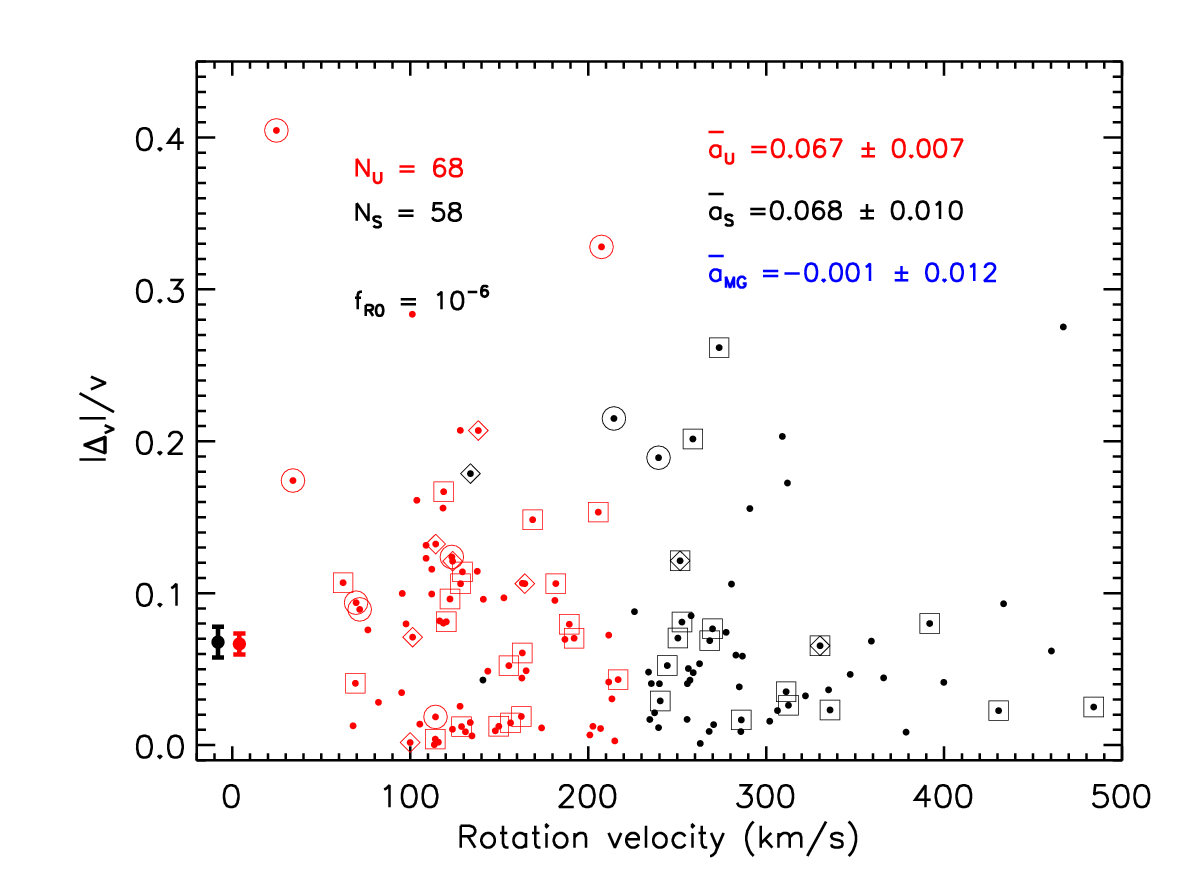}
\caption{The  \ha{} rotation curve asymmetry $a=|\Delta v|/v$ is shown as a
function of rotation velocity for  galaxies from the GHASP survey. Black (red) points
indicate screened (unscreened) galaxies for $f_{R0}=10^{-6}$.
The mean and standard deviations of each distribution are shown by the large
dots at the left of the plot. The shapes of individual points indicate the 
morphology of the rotation curves: diamonds indicate smoothly asymmetric
curves, squares indicate barred galaxies, and empty circles indicate irregular
galaxies.  See  text for details.
\label{fig:plotsGHASP}}
\end{center}
\end{figure}

In the previous section, we show that if MG is true, then \ha{} empirically
traces the unscreened gaseous component of the galaxy. This implies
that the \ha{} rotation curve will be nearly symmetric about the
center of dark matter halo. On the other hand, the dark matter halo will be
offset from the stellar component and therefore the \ha{} curve will be
asymmetric with respect to the optical centroid, with an asymmetry on
the order of 10-15km/s. We note that this asymmetry is
a manifestation of the offset between the stellar and dark matter halo,
and is distinct from the stellar rotation curve asymmetry predicted by
\citet{bhuvjake2011}.

In this section we use \ha{} rotation curves to test
whether gaseous rotation curves are asymmetric about the stellar centroid,
by comparing the approaching and receding sides of the \ha{} rotation curves
about the associated stellar centroid. Any
positive signal may indicate the presence of fifth force.

\subsubsection{Data and visual classification of rotation curves}
We use \ha{} rotation curves for nearly 200 disk
galaxies from  the Gassendi \ha{} survey of spirals (GHASP) archive
\citep[see][and references therein]{epinat08a} to search for asymmetry in
rotation curves. GHASP is an \ha{} kinematic survey of spiral and
irregular galaxies. This survey has observed $\sim 220$ galaxies, and
provides the rotation curves for both
approaching and receding sides separately. We correct the velocities for
inclination by assuming thin circular disk.

We visually inspect these rotation curves and classify them into symmetric,
smoothly asymmetric and irregular. We show a sample of symmetric curves in
Figure
\ref{fig:GHASPsym} and smoothly asymmetric in Figure \ref{fig:GHASPasym} in
Appendix \ref{ap:curves}. In a modified gravity scenario, we expect the gaseous
rotation curves to be smoothly asymmetric about the stellar
centroid \citep{bhuvjake2011}.
We remove all the irregular rotation curves which have too few
observed points and/or small-scale discontinuities: these are primarily
associated with small, dim and/or morphologically irregular galaxies, and the
rotation curve shapes are more likely due to small-scale astrophysical
effects than to a fifth-force type effect.
We find that in both the screened and unscreened sample, symmetric curves
outnumber smoothly asymmetric curves by a factor of four.
We emphasize, however, that the galaxies with asymmetric \ha{} curves may
be ideal for a future systematic search for the dynamical effects of
modified gravity.

\subsubsection{Analysis}
We quantify the asymmetry in rotation curves for each galaxy
using the abolute value of the weighted average of difference between
approaching and receding sides of the rotation curves ($|\Delta v|$) about the
stellar centroids \citep{epinat08a}. This asymmetry is better seen near the
center of the galaxy and therefore we focus on the inner part of the
rotation curves. However, we find that our results do not depend
significantly on the location of the measurement. Finally, we define the
asymmetry coefficient $a \equiv |\Delta v| / |v_{max}|$ where $v_{max}$ is the
maximum rotation velocity, since we expect any asymmetries to scale with
galaxy mass.
The error in the asymmetry
includes both measurement error and intrinsic dispersion $\sigma_a$, which
can be same order of magnitude; we follow the method described in Appendix
\ref{ap:avg}. In a similar manner as the previous sections, we assume that
the asymmetry of unscreened galaxy, $a^U$ is a linear combination of
$a^{MG}$, the modified gravity contribution, and $a^X$, the contribution from
standard astrophysical forces.

%

\subsubsection{Results}
We summarize the results in Figure \ref{fig:plotsGHASP}.  We plot  $a$
as a function of maximum rotation velocity of the galaxies for
$f_{R0}=10^{-6}$. There are 68 unscreened galaxies
(shown in red) and 58 screened galaxies (shown in black).
We compute the asymmetry within the central region of the rotation curve,
up to 30\% the optical radius of the galaxy.

The shapes of the individual symbols indicate the visually-determined galaxy
morphology: the diamond symbols indicate
galaxies with visual asymmetry in the rotation curve which look interesting
under modified gravity theories. Square symbols indicate barred
galaxies and empty circles represent irregular cases.
There is no clear correlation between these morphologies and the
measured asymmetry.   The filled points on the left of the plot show the
mean and standard deviation of the asymmetries for the screened and unscreened
components. We find that the asymmetry for both
screened and unscreened galaxies is around 7\% which corresponds to
asymmetries in the range of 10-20 km/s.
Subtracting these mean values implies an asymmetry produced by modified
gravity of $a^{MG}=-0.001\pm 0.012$.
If we instead assume $f_{R0}=10^{-6.5}$ (24 unscreened galaxies and 102
screened galaxies) we find $a^{MG}=0.024\pm 0.015$.
In either case, a 10\% asymmetry predicted for modified gravity can be
ruled out at approximately the 5-$\sigma$ level.

Note that
these results do not attempt to correct for asymmetric drift or non-circular
motion.  We have further analyzed this data and found no correlation between
asymmetry and the number of neighbors or distance to the nearest neighbor,
which would be present if the asymmetries were caused by a fifth force.
\citet{garrido05} also studied asymmetry in
GHASP \ha{} rotational curves. They conclude that both isolated and softly
interacting galaxies have similar level of asymmetry, which might be due to
inhomogeneous massive star formation regions.

\section{Discussion and Conclusions}

\label{sec:discussion}
We have explored several astrophysical tests of modified gravity which rely on
partially-screened dwarf galaxies in which fifth force effects act on
diffuse components (gas, dark matter, and giant stars), while more
compact components (main sequence stars) are screened from these effects.
These tests probe regions of the parameter space of chameleon type gravity
theories which have have not been probed by cosmological tests or
laboratory experiments.  The observables include offsets between the stellar
and gaseous components, warping of stellar disks, and asymmetry in the
gaseous rotation curves. Our study relies on a variety of previously-published
observational data.

\subsection{Summary of results}

\begin{itemize}
\item {\it Offset between HI gas and stars:}
The offset study based on optical and HI gas centroids shows that it is not
possible to distinguish between gravity theories using the current HI data. 
The accuracy of the astrometry is principally limited by the large beam size of
the 21cm observations.  However, with a larger catalog of candidate
galaxies and/or more accurate radio astrometry, this test may provide
competitive constraints in the future. In this section we summarize our results, 
discuss systematic errors and examine prospects for improved tests. 

\item {\it Offset between optical and dynamical center:}
An alternative means of measuring this expected offset, the difference
between the optical centroid and the kinematical centroid of \ha{}, is
much more precise and limits the component of modified gravity scatter
to  $\sim 2$\% of the optical radius of the galaxy, ruling out 5\%
scatter at the 3-$\sigma$ level.

\item {\it Offset between red giants and main sequence stars:}
We show that the offset between the partially-screened red giant
population and the fully-screened main sequence population is too
sensitive by systematic effects to provide any meaningful power to distinguish
between gravity theories: the data is consistent with both GR and MG at
a 2-$\sigma$ level.  This test has the potential to improve as more stellar
populations are observed within nearby dwarf galaxies.

\item {\it Warps in edge-on galaxies:}
The warp study relies on measurements of the optical morphology of stellar
disks of late-type galaxies observed by SDSS.
We find that the warp measures for this galaxy sample are
consistent with both GR and modified gravity theories: the main limitation
is the relatively small number of suitable galaxies for which kinematic data
is available in order to estimate the screening level of the galaxies in
the sample.   

\item {\it Rotation curve tests: }
In the final test we used the GHASP galaxy sample to estimate
the asymmetry in the \ha{} rotation curves, which arises due to the
offset between the screened stellar component and the unscreened gaseous
component. We find that the asymmetry attributable to modified gravity is as
low as (-0.1 $\pm$ 1)\%, compared to the expected value of 10\% under
favorable conditions.
This implies that the modified gravity models which predicts 10\%
asymmetry can be rule out at $5 \sigma$ level. However, this test carries two
major uncertainties: systematic uncertainties in the data due to effects of
asymmetric drift and non-circular motions, and systematic uncertainties in
the theoretical expectation due to insufficient knowledge of the internal
halo parameters and external environment of each galaxy.
\end{itemize}

Finally, we examined individual rotation curves from the GHASP 
galaxy sample and classified them based on the different signatures 
in the curves (see Appendix D). We found that the majority of these galaxies 
have fairly symmetric curves regardless
of their local density/potential. However, we identified a 
few galaxies with interesting features in the rotation curves.
Some of these galaxies have other potential fifth force effects
like U-type warps and optical offset from that of HI.
More detailed studies of these individual cases can shed light
on whether the features may be attributable to fifth-force effects or
more traditional astrophysical processes.

\subsection{Systematic errors and caveats}
\label{sec:caveat}
The  tests performed here involve aggregate statistical
comparisons between the unscreened galaxy sample and the screened (control) sample.
One of the major assumptions in this type of study is that, apart
from screening level, the two samples are statistically similar.  As noted
above, for many of the tests this division correlates with galaxy size. 
In some cases we attempt to minimize any potential bias
by looking only at galaxies near the self-screening threshold (at the cost
of weakening the signal).  However,
a more precise result could be obtained given a large sample of low-mass,
environmentally screened galaxies.  Galaxies in the two samples would then
have the same distribution in size and mass. So we recommend that
future studies in this direction prioritize new observations of low-mass
galaxies in both the field and near large neighbors.

Another important caveat is that we use a large, heterogenous data set to
perform the above tests.  Because the data were gathered without this
type of test in mind, there was little specific attempt to control
systematic errors that can affect explorations of modified gravity.
One extreme example of this is the constraint based on separation of
HI and optical centroids: the strength of the test is sensitive to the
 characteristics of the error in centroid determination: even a
10\% overestimate of the error can impact the
resulting constraints.  Observing campaigns specifically designed for this
sort of modified gravity test could yield homogeneous data sets with such
systematic effects minimized.

The third major caveat comes from uncertainties in our estimate of the 
external potential at the locations of dwarf galaxies. We have estimated the 
Newtonian potential as well as the fifth force, estimated using the masses 
and locations of neighboring galaxies. This estimate is subject to the 
uncertainties in mass estimation of galaxies. It can be improved in a number of 
ways: completeness over the neighborhood of the dwarf galaxy
sample, mass estimates using spectroscopic or 21cm observations, and better modeling of the mass and scalar field distribution. The improved modeling 
can be done by explicitly solving for the scalar field at each location, including the 
presence of underdense regions that also contribute to the fifth force. For some
of the tests only a $\sim 10$ Mpc sized region needs to studied in detail, so 
the observations and modeling are tractable. 

\subsection{Future prospects}
Here we discuss specific ways to sharpen each of the tests we have carried out and summarized above in subsection 7.1. Before we get to these tests, we note that a variety of experiments can be used to test gravity theories, ranging from laboratory to cosmological scales. For chameleon type scalar-tensor theories, the first in our series of papers \citep{jainvinu2012} used cepheid variable stars and other distance indicators to place stringent limits on the two parameters of the theories: the coupling $\alpha$ (related to $\Delta G/G$ in this paper) and the background field value (equivalent to $f_{R0}$ in this paper).  The upper limit for $f(R)$ models obtained using cepheids correspond to $\Delta G/G\simeq 1/3; f_{R0}\simeq 5\times10^{-7}$, or a fifth force range below 1 Mpc at cosmic mean density: these are far more stringent than limits from large-scale, cosmological tests as discussed by \citet{jainvinu2012}. For tests using other tracers in the Milky Way, see \citet{Brax2013} and reference therein. 

Dwarf galaxy tests currently probe larger coupling values. However the field values probed can be lower, especially with the fainter dwarf galaxies within $\sim 10$ Mpc of the Milky Way. The limits on both parameters can be improved significantly 
for all the tests we have considered. A brief account for each test follows, though 
detailed studies are needed to plan future observations.  

\begin{itemize}
\item {\it Offset between HI gas and stars:}
  For this test to produce competitive constraints, it is crucial to improve
  the precision of the centroid measurement, either through telescopes with
  a smaller beam size, or through the stacking of many more galaxy images.
  Using the ALFALFA data, the error in the offset estimate is $\sim$ 1 kpc;
  to get an error of 100 pc we need to stack $\sim 10^{4}$
  galaxies.  Furthermore, the characteristics of the error
  in the centroid must be understood at the few percent level. A better way forward
  would be to get improved resolution radio maps (e.g. via VLA observations), 
  which would then relax
  the required sample size and lower the systematic uncertainty. 
\item {\it Offset between optical and dynamical center:}
  increasing the number of \ha~ rotation curves to a few hundred 
  low-mass galaxies would be sufficient to reduce the uncertainty of
  this measurement.  More important is sophisticated
  modeling of the environment of each galaxy to determine screening levels and
  the strength of the expected signal.
\item {\it Offset between red giants and main sequence stars:}
  Currently, the red giant displacement test is dominated primarily by
  systematic uncertainties. As the signal in this test is weaker than the
  other probes, the best approach would be to focus on a limited number
  of galaxies with a very well known external force (i.e. well-observed and
  well-modeled local environment). The uncertainty in the external force is the
  limiting factor if one wants to constrain modified gravity.  Another potential
  approach would be to use integral-field galaxy spectra and estimate the
  contributions of red giant and main sequence populations as a function of
  location in the galaxy.  A difference in these centroids could be
  attributed to modified gravity, using control tests similar to those in
  this paper.  This test is applied to a small number of 
  very small but nearby galaxies (closer than 10 Mpc), so detailed studies of individual 
  galaxies are feasible. 
\item {\it Warps in edge-on galaxies:}
  To reduce the observed error in the warp measurements 
  to the level of uncertainty in the 
  predicted values requires high-quality images
  of about 10,000 dwarf galaxies in very low density environments and 20,000
  dwarf galaxies in low to moderately dense environments for $f(R)$ theories
  with $f_{R0} = 2\times10^{-7}$ and $f_{R0} = 1\times10^{-6}$, respectively.
  These numbers are feasible with  current and ongoing galaxy surveys.
  With this increase in data, however, the limiting factor will be the
  knowledge of the distribution of halo parameters within galaxies of this
  size: the predicted strength of the observed warp is highly dependent on
  this mass profile. Rotation curves that extend to the inner parts of the disk
can
  constrain the mass profile. 
\item {\it  Rotation curve tests: }
  We have examined asymmetries in rotation curve data. In addition, 
  a direct measurement of $\Delta G/G$ can be made by
  comparing HI rotation curves with stellar rotation curves (section 3.1). However 
  the stellar rotation curve must be measured using stellar absorption
  lines which arise very close to the surface of dwarf stars (not via 
  \ha{} as is common practice; see discussion in section 3.2).  These curves
  are difficult to observe: depending on brightness, each galaxy may require
  several hours of observations with large telescopes. This test
  has the advantage that it does not depend on a detailed dynamical model
  or on the precise characteristics of the environment of the unscreened
  galaxy: for this reason, high-precision observations of
  a handful of the unscreened galaxies can lead to strong constraints on
  modified gravity theories, free of many of the systematic uncertainties
  discussed above.
\end{itemize}

\section{Acknowledgments}
We are grateful to Xiaohu Yang and collaborators 
for providing us their SDSS group catalog and to Richard Swaters for 
rotation curve data. We have used published data from a variety of sources
cited in the text.  
We thank Mike Jarvis for several valuable suggestions and Gary 
Bernstein, Andy Connolly, Julianne Dalcanton, Lam Hui, Kazuya Koyama, 
Elisabeth Krause, Adam Lidz, Karen Masters, Masao Sako, 
Matt Walker and Gongbo Zhao for helpful
discussions. This work was partially supported by NSF grant AST-0908027.

Funding for the SDSS and SDSS-II has been provided by the Alfred P. Sloan
Foundation, the Participating Institutions, the National Science Foundation,
the U.S. Department of Energy, the National Aeronautics and Space
Administration, the Japanese Monbukagakusho, the Max Planck Society, and the
Higher Education Funding Council for England. The SDSS Web Site is
http://www.sdss.org/.

The SDSS is managed by the Astrophysical Research Consortium for the
Participating Institutions. The Participating Institutions are the American
Museum of Natural History, Astrophysical Institute Potsdam, University of
Basel, University of Cambridge, Case Western Reserve University, University of
Chicago, Drexel University, Fermilab, the Institute for Advanced Study, the
Japan Participation Group, Johns Hopkins University, the Joint Institute for
Nuclear Astrophysics, the Kavli Institute for Particle Astrophysics and
Cosmology, the Korean Scientist Group, the Chinese Academy of Sciences
(LAMOST), Los Alamos National Laboratory, the Max-Planck-Institute for
Astronomy (MPIA), the Max-Planck-Institute for Astrophysics (MPA), New Mexico
State University, Ohio State University, University of Pittsburgh,
University of Portsmouth, Princeton University, the United States Naval
Observatory, and the University of Washington.

\bibliographystyle{mn2e}
\bibliography{dwarfdraft}

\begin{thebibliography}{}

\bibitem[\protect\citeauthoryear{{Abazajian}, {Adelman-McCarthy},
  {Ag{\"u}eros}, {Allam}, {Allende Prieto}, {An}, {Anderson}, {Anderson},
  {Annis}, {Bahcall} \& et al.}{{Abazajian} et~al.}{2009}]{aba09}
{Abazajian} K.~N.,  {Adelman-McCarthy} J.~K.,  {Ag{\"u}eros} M.~A.,  {Allam}
  S.~S.,  {Allende Prieto} C.,  {An} D.,  {Anderson} K.~S.~J.,  {Anderson}
  S.~F.,  {Annis} J.,  {Bahcall} N.~A.,    et al. 2009, \apjs, 182, 543

\bibitem[\protect\citeauthoryear{{Abell}, {Corwin} Jr. \& {Olowin}}{{Abell}
  et~al.}{1989}]{abe89}
{Abell} G.~O.,  {Corwin} Jr. H.~G.,    {Olowin} R.~P.,  1989, \apjs, 70, 1

\bibitem[\protect\citeauthoryear{{Begum}, {Chengalur}, {Karachentsev}, {Kaisin}
  \& {Sharina}}{{Begum} et~al.}{2006}]{begum06}
{Begum} A.,  {Chengalur} J.~N.,  {Karachentsev} I.~D.,  {Kaisin} S.~S.,
  {Sharina} M.~E.,  2006, \mnras, 365, 1220

\bibitem[\protect\citeauthoryear{{Begum}, {Chengalur}, {Karachentsev},
  {Sharina} \& {Kaisin}}{{Begum} et~al.}{2008}]{figgs2008}
{Begum} A.,  {Chengalur} J.~N.,  {Karachentsev} I.~D.,  {Sharina} M.~E.,
  {Kaisin} S.~S.,  2008, \mnras, 386, 1667

\bibitem[\protect\citeauthoryear{{Bertin} \& {Arnouts}}{{Bertin} \&
  {Arnouts}}{1996}]{bertin96}
{Bertin} E.,  {Arnouts} S.,  1996, \aaps, 117, 393

\bibitem[\protect\citeauthoryear{{Brax} \& {Davis}}{{Brax} \&
  {Davis}}{2013}]{Brax2013}
{Brax} P.,  {Davis} A.-C.,  2013, ArXiv e-prints

\bibitem[\protect\citeauthoryear{{Brax}, {van de Bruck}, {Davis} \&
  {Shaw}}{{Brax} et~al.}{2010}]{brax10}
{Brax} P.,  {van de Bruck} C.,  {Davis} A.-C.,    {Shaw} D.,  2010, \prd, 82,
  063519

\bibitem[\protect\citeauthoryear{{Cabr{\'e}}, {Vikram}, {Zhao}, {Jain} \&
  {Koyama}}{{Cabr{\'e}} et~al.}{2012}]{cabre2012}
{Cabr{\'e}} A.,  {Vikram} V.,  {Zhao} G.-B.,  {Jain} B.,    {Koyama} K.,  2012,
  ArXiv e-prints

\bibitem[\protect\citeauthoryear{{Chang} \& {Hui}}{{Chang} \&
  {Hui}}{2011}]{changhui}
{Chang} P.,  {Hui} L.,  2011, \apj, 732, 25

\bibitem[\protect\citeauthoryear{{Dalcanton}, {Williams}, {Seth}, {Dolphin},
  {Holtzman}, {Rosema}, {Skillman}, {Cole}, {Girardi}, {Gogarten} \&
  {Karachentsev}}{{Dalcanton} et~al.}{2009}]{dalcanton09}
{Dalcanton} J.~J.,  {Williams} B.~F.,  {Seth} A.~C.,  {Dolphin} A.,  {Holtzman}
  J.,  {Rosema} K.,  {Skillman} E.~D.,  {Cole} A.,  {Girardi} L.,  {Gogarten}
  S.~M.,    {Karachentsev} I.~D.,  2009, \apjs, 183, 67

\bibitem[\protect\citeauthoryear{{Davis}, {Lim}, {Sakstein} \& {Shaw}}{{Davis}
  et~al.}{2012}]{Davis2012}
{Davis} A.-C.,  {Lim} E.~A.,  {Sakstein} J.,    {Shaw} D.~J.,  2012, \prd, 85,
  123006

\bibitem[\protect\citeauthoryear{{de Blok} \& {Bosma}}{{de Blok} \&
  {Bosma}}{2002}]{deblok2002}
{de Blok} W.~J.~G.,  {Bosma} A.,  2002, \aap, 385, 816

\bibitem[\protect\citeauthoryear{{Ebeling}, {Voges}, {Bohringer}, {Edge},
  {Huchra} \& {Briel}}{{Ebeling} et~al.}{1996}]{ebe96}
{Ebeling} H.,  {Voges} W.,  {Bohringer} H.,  {Edge} A.~C.,  {Huchra} J.~P.,
  {Briel} U.~G.,  1996, \mnras, 281, 799

\bibitem[\protect\citeauthoryear{{Epinat}, {Amram}, {Marcelin}, {Balkowski},
  {Daigle}, {Hernandez}, {Chemin}, {Carignan}, {Gach} \& {Balard}}{{Epinat}
  et~al.}{2008}]{epinat08a}
{Epinat} B.,  {Amram} P.,  {Marcelin} M.,  {Balkowski} C.,  {Daigle} O.,
  {Hernandez} O.,  {Chemin} L.,  {Carignan} C.,  {Gach} J.-L.,    {Balard} P.,
  2008, \mnras, 388, 500

\bibitem[\protect\citeauthoryear{{Evrard}, {Bialek}, {Busha}, {White}, {Habib},
  {Heitmann}, {Warren}, {Rasia}, {Tormen}, {Moscardini}, {Power}, {Jenkins},
  {Gao}, {Frenk}, {Springel}, {White} \& {Diemand}}{{Evrard}
  et~al.}{2008}]{evr08}
{Evrard} A.~E.,  {Bialek} J.,  {Busha} M.,  {White} M.,  {Habib} S.,
  {Heitmann} K.,  {Warren} M.,  {Rasia} E.,  {Tormen} G.,  {Moscardini} L.,
  {Power} C.,  {Jenkins} A.~R.,  {Gao} L.,  {Frenk} C.~S.,  {Springel} V.,
  {White} S.~D.~M.,    {Diemand} J.,  2008, \apj, 672, 122

\bibitem[\protect\citeauthoryear{{Garrido}, {Marcelin}, {Amram}, {Balkowski},
  {Gach} \& {Boulesteix}}{{Garrido} et~al.}{2005}]{garrido05}
{Garrido} O.,  {Marcelin} M.,  {Amram} P.,  {Balkowski} C.,  {Gach} J.~L.,
  {Boulesteix} J.,  2005, \mnras, 362, 127

\bibitem[\protect\citeauthoryear{{Geha}, {Blanton}, {Masjedi} \& {West}}{{Geha}
  et~al.}{2006}]{geha06}
{Geha} M.,  {Blanton} M.~R.,  {Masjedi} M.,    {West} A.~A.,  2006, \apj, 653,
  240

\bibitem[\protect\citeauthoryear{{Giovanelli}, {Haynes}, {Kent}, {Perillat},
  {Saintonge}, {Brosch}, {Catinella}, {Hoffman}, {Stierwalt}, {Spekkens},
  {Lerner} \& {Masters}}{{Giovanelli} et~al.}{2005}]{giovanelli2005}
{Giovanelli} R.,  {Haynes} M.~P.,  {Kent} B.~R.,  {Perillat} P.,  {Saintonge}
  A.,  {Brosch} N.,  {Catinella} B.,  {Hoffman} G.~L.,  {Stierwalt} S.,
  {Spekkens} K.,  {Lerner} M.~S.,    {Masters} K.~L.,  2005, \aj, 130, 2598

\bibitem[\protect\citeauthoryear{{Giovanelli}, {Haynes}, {Kent}, {Saintonge},
  {Stierwalt}, {Altaf}, {Balonek}, {Brosch}, {Brown}, {Catinella}, {Furniss} \&
  {Goldstein}}{{Giovanelli} et~al.}{2007}]{gio07}
{Giovanelli} R.,  {Haynes} M.~P.,  {Kent} B.~R.,  {Saintonge} A.,  {Stierwalt}
  S.,  {Altaf} A.,  {Balonek} T.,  {Brosch} N.,  {Brown} S.,  {Catinella} B.,
  {Furniss} A.,    {Goldstein} J.,  2007, \aj, 133, 2569

\bibitem[\protect\citeauthoryear{{Hinterbichler} \& {Khoury}}{{Hinterbichler}
  \& {Khoury}}{2010}]{hinterbichler10}
{Hinterbichler} K.,  {Khoury} J.,  2010, Physical Review Letters, 104, 231301

\bibitem[\protect\citeauthoryear{{Hu} \& {Sawicki}}{{Hu} \&
  {Sawicki}}{2007}]{hu07}
{Hu} W.,  {Sawicki} I.,  2007, \prd, 76, 064004

\bibitem[\protect\citeauthoryear{{Hui} \& {Nicolis}}{{Hui} \&
  {Nicolis}}{2010}]{hui10}
{Hui} L.,  {Nicolis} A.,  2010, Physical Review Letters, 105, 231101

\bibitem[\protect\citeauthoryear{{Hui}, {Nicolis} \& {Stubbs}}{{Hui}
  et~al.}{2009}]{hui09}
{Hui} L.,  {Nicolis} A.,    {Stubbs} C.~W.,  2009, \prd, 80, 104002

\bibitem[\protect\citeauthoryear{{Jain}}{{Jain}}{2011}]{jain11}
{Jain} B.,  2011, Royal Society of London Philosophical Transactions Series A,
  369, 5081

\bibitem[\protect\citeauthoryear{{Jain} \& {Khoury}}{{Jain} \&
  {Khoury}}{2010}]{jain-khoury10}
{Jain} B.,  {Khoury} J.,  2010, Annals of Physics, 325, 1479

\bibitem[\protect\citeauthoryear{{Jain} \& {Vanderplas}}{{Jain} \&
  {Vanderplas}}{2011}]{bhuvjake2011}
{Jain} B.,  {Vanderplas} J.,  2011, ArXiv e-prints

\bibitem[\protect\citeauthoryear{{Jain}, {Vikram} \& {Sakstein}}{{Jain}
  et~al.}{2012}]{jainvinu2012}
{Jain} B.,  {Vikram} V.,    {Sakstein} J.,  2012, ArXiv e-prints

\bibitem[\protect\citeauthoryear{{Jim{\'e}nez-Vicente}, {Porcel},
  {S{\'a}nchez-Saavedra} \& {Battaner}}{{Jim{\'e}nez-Vicente}
  et~al.}{1997}]{jim97}
{Jim{\'e}nez-Vicente} J.,  {Porcel} C.,  {S{\'a}nchez-Saavedra} M.~L.,
  {Battaner} E.,  1997, \apss, 253, 225

\bibitem[\protect\citeauthoryear{{Karachentsev}, {Karachentseva}, {Huchtmeier}
  \& {Makarov}}{{Karachentsev} et~al.}{2004}]{kar04}
{Karachentsev} I.~D.,  {Karachentseva} V.~E.,  {Huchtmeier} W.~K.,    {Makarov}
  D.~I.,  2004, \aj, 127, 2031

\bibitem[\protect\citeauthoryear{{Kent}, {Giovanelli}, {Haynes}, {Martin},
  {Saintonge}, {Stierwalt}, {Balonek}, {Brosch} \& {Koopmann}}{{Kent}
  et~al.}{2008}]{ken08}
{Kent} B.~R.,  {Giovanelli} R.,  {Haynes} M.~P.,  {Martin} A.~M.,  {Saintonge}
  A.,  {Stierwalt} S.,  {Balonek} T.~J.,  {Brosch} N.,    {Koopmann} R.~A.,
  2008, \aj, 136, 713

\bibitem[\protect\citeauthoryear{{Khoury} \& {Weltman}}{{Khoury} \&
  {Weltman}}{2004}]{khoury04}
{Khoury} J.,  {Weltman} A.,  2004, \prd, 69, 044026

\bibitem[\protect\citeauthoryear{{Lavaux} \& {Hudson}}{{Lavaux} \&
  {Hudson}}{2011}]{lav011}
{Lavaux} G.,  {Hudson} M.~J.,  2011, \mnras, 416, 2840

\bibitem[\protect\citeauthoryear{{Martin}, {Giovanelli}, {Haynes}, {Saintonge},
  {Hoffman}, {Kent} \& {Stierwalt}}{{Martin} et~al.}{2009}]{mar09}
{Martin} A.~M.,  {Giovanelli} R.,  {Haynes} M.~P.,  {Saintonge} A.,  {Hoffman}
  G.~L.,  {Kent} B.~R.,    {Stierwalt} S.,  2009, \apjs, 183, 214

\bibitem[\protect\citeauthoryear{{Mathewson}, {Ford} \& {Buchhorn}}{{Mathewson}
  et~al.}{1992}]{mat92}
{Mathewson} D.~S.,  {Ford} V.~L.,    {Buchhorn} M.,  1992, \apjs, 81, 413

\bibitem[\protect\citeauthoryear{{Persic} \& {Salucci}}{{Persic} \&
  {Salucci}}{1995}]{persic1995}
{Persic} M.,  {Salucci} P.,  1995, \apjs, 99, 501

\bibitem[\protect\citeauthoryear{{Reiprich} \& {B{\"o}hringer}}{{Reiprich} \&
  {B{\"o}hringer}}{2002}]{rei02}
{Reiprich} T.~H.,  {B{\"o}hringer} H.,  2002, \apj, 567, 716

\bibitem[\protect\citeauthoryear{{Saintonge}, {Giovanelli}, {Haynes},
  {Hoffman}, {Kent}, {Martin}, {Stierwalt} \& {Brosch}}{{Saintonge}
  et~al.}{2008}]{sai08}
{Saintonge} A.,  {Giovanelli} R.,  {Haynes} M.~P.,  {Hoffman} G.~L.,  {Kent}
  B.~R.,  {Martin} A.~M.,  {Stierwalt} S.,    {Brosch} N.,  2008, \aj, 135, 588

\bibitem[\protect\citeauthoryear{{Schmidt}, {Vikhlinin} \& {Hu}}{{Schmidt}
  et~al.}{2009}]{sch09}
{Schmidt} F.,  {Vikhlinin} A.,    {Hu} W.,  2009, \prd, 80, 083505

\bibitem[\protect\citeauthoryear{{Swaters}, {Madore}, {van den Bosch} \&
  {Balcells}}{{Swaters} et~al.}{2003}]{swaters03}
{Swaters} R.~A.,  {Madore} B.~F.,  {van den Bosch} F.~C.,    {Balcells} M.,
  2003, \apj, 583, 732

\bibitem[\protect\citeauthoryear{{Swaters}, {Sancisi}, {van Albada} \& {van der
  Hulst}}{{Swaters} et~al.}{2009}]{Swaters2009}
{Swaters} R.~A.,  {Sancisi} R.,  {van Albada} T.~S.,    {van der Hulst} J.~M.,
  2009, \aap, 493, 871

\bibitem[\protect\citeauthoryear{{Swaters}, {Sancisi}, {van Albada} \& {van der
  Hulst}}{{Swaters} et~al.}{2011}]{swaters11}
{Swaters} R.~A.,  {Sancisi} R.,  {van Albada} T.~S.,    {van der Hulst} J.~M.,
  2011, \apj, 729, 118

\bibitem[\protect\citeauthoryear{{Swaters}, {van Albada}, {van der Hulst} \&
  {Sancisi}}{{Swaters} et~al.}{2002}]{swaters2002b}
{Swaters} R.~A.,  {van Albada} T.~S.,  {van der Hulst} J.~M.,    {Sancisi} R.,
  2002, \aap, 390, 829

\bibitem[\protect\citeauthoryear{{Tully} \& {Fouque}}{{Tully} \&
  {Fouque}}{1985}]{tully85}
{Tully} R.~B.,  {Fouque} P.,  1985, \apjs, 58, 67

\bibitem[\protect\citeauthoryear{{Will}}{{Will}}{2006}]{will06}
{Will} C.~M.,  2006, Living Reviews in Relativity, 9, 3

\bibitem[\protect\citeauthoryear{{Yang}, {Mo}, {van den Bosch}, {Pasquali},
  {Li} \& {Barden}}{{Yang} et~al.}{2007}]{yan07}
{Yang} X.,  {Mo} H.~J.,  {van den Bosch} F.~C.,  {Pasquali} A.,  {Li} C.,
  {Barden} M.,  2007, \apj, 671, 153

\end{thebibliography}

\clearpage

\appendix

\section{Algorithm to find intrinsic dispersion}
\label{ap:avg}

In most of the probes that we consider in this paper, there is an intrinsic
dispersion associated with the estimated parameter due to astrophysical
systematic effects that can be
larger than the error in individual measurements.
Because we weight the data by the inverse variance while calculating the
mean value, including an intrinsic dispersion keeps data points 
with small measurement errors from being inappropriately over-weighted. It 
leads to an improved  estimate of the mean and the error in the mean value.
 Here we describe our approximate procedure to estimate the intrinsic
dispersion,  and the improved mean and the error in the mean.


We have set of N observed values $x_i$ and errors $\sigma_i$. We calculate the
weighted average as 
\begin{equation}
 \bar{x} = \frac{\sum_i{x_i w_i}}{\sum_i{w_i}}
\label{eq:weightavg}
\end{equation}
where $w_i = 1/\sigma_i^2$. Now, we add a constant systematic error $\sigma_s$
to all the
individual errors as $\sigma_{{\rm eff},i}^2 = \sigma_i^2 + \sigma_s^2$ which
accounts for the intrinsic dispersion. 
For a Gaussian random variable this satisfies:
\begin{equation}
 \sum_i{\frac{(x_i - \bar{x})^2}{\sigma_{{\rm eff},i}^2}} = N
\end{equation}
Now we recalculate the mean $\bar{x}$ using Eq. \ref{eq:weightavg} with $w_i =
1/\sigma_{{\rm eff},i}^2$. This process continues until it converges. Finally, the
error on the average $\bar{x}$ will be calculated by
\begin{equation}
 \sigma_{\bar{x}}^2 = \frac{1}{\sum_i{1 / \sigma_{{\rm eff},i}^2}}
\end{equation}

\section{Red giant stars}
\label{ap:rgb-systematics}

\begin{figure}
\centering{
\includegraphics[trim= 2cm 25cm 15cm 10cm,  scale=0.3, clip =
true]{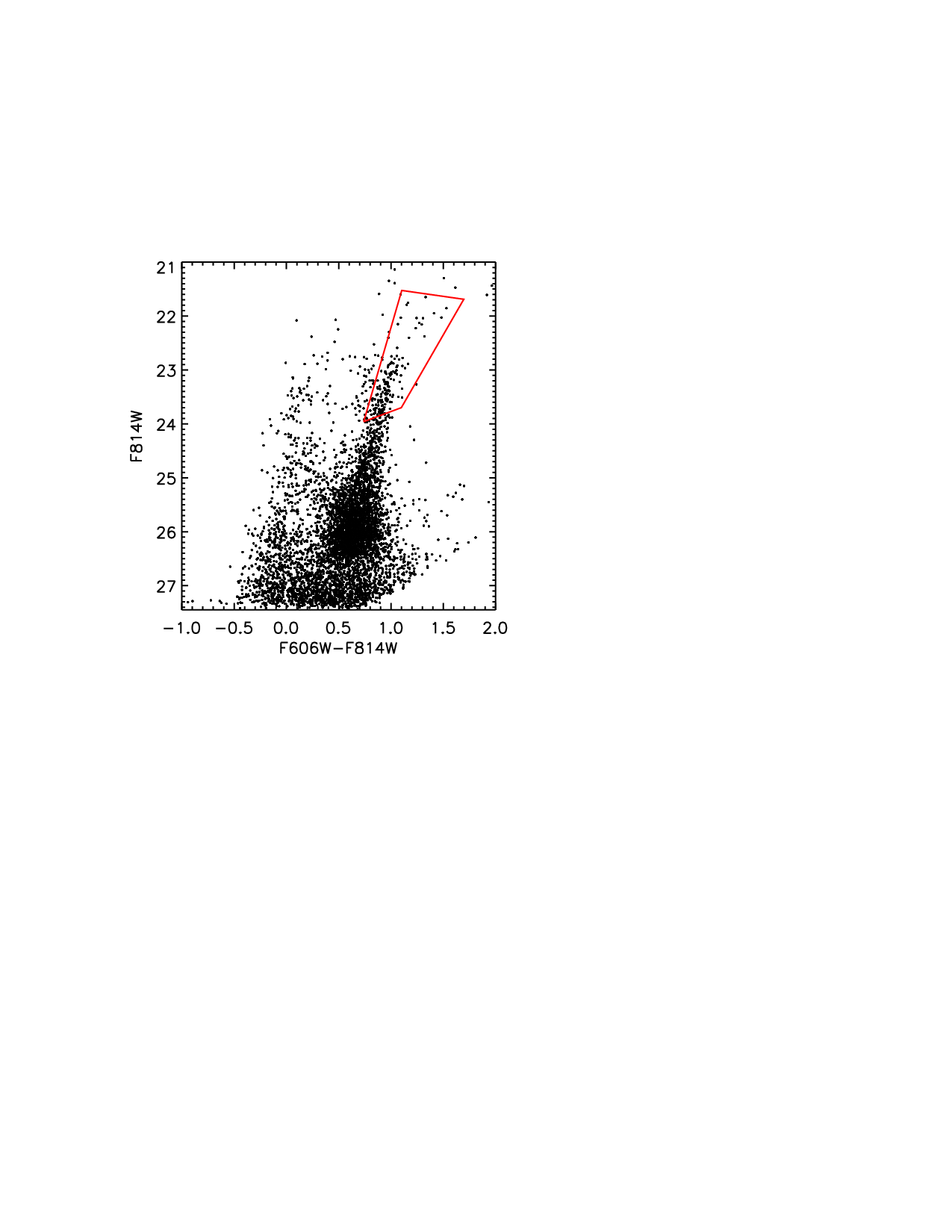}}
  \caption{Color-magnitude diagram for galaxy DDO187, using filters F606W
and F814W. The red region shows the selected red giants, that reside at the
bright tip of the red giant branch. \label{fig:selectrg}}
\end{figure}

In this section we summarise the potential systematic errors related to test
based on red giant stars.
In Figure \ref{fig:selectrg} we show an example of color-magnitude diagram to
illustrate the selection of red giant stars. 

\begin{figure*}
\includegraphics[scale=0.2]{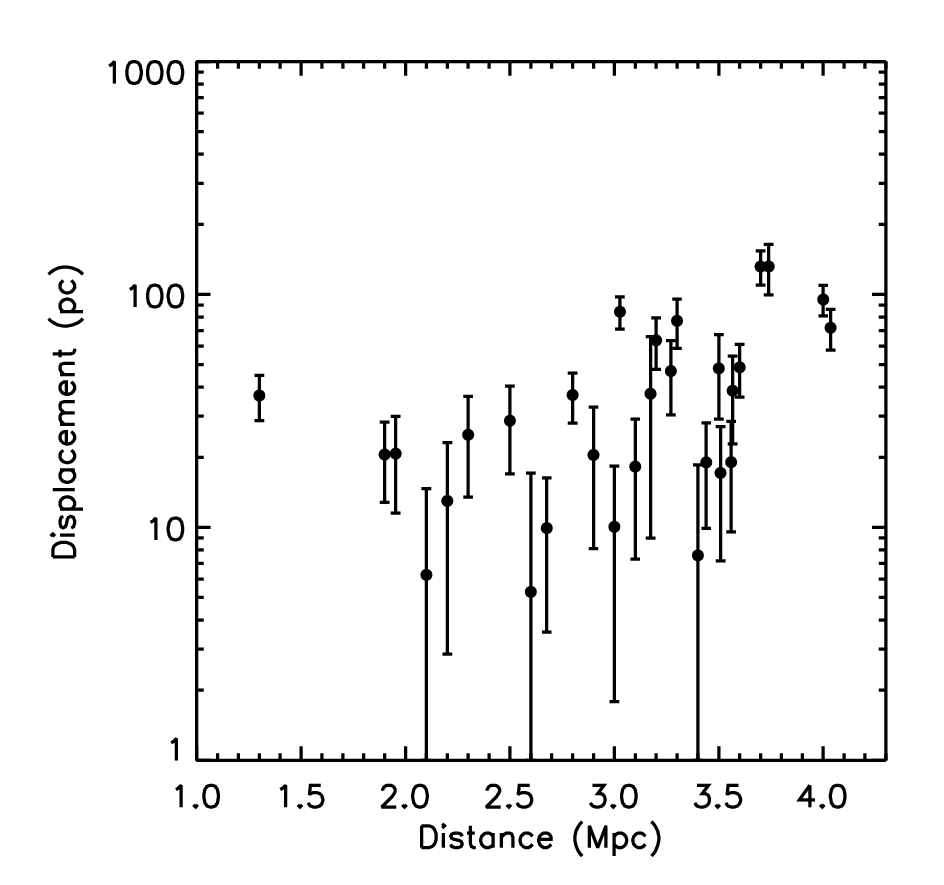}
\includegraphics[scale=0.2]{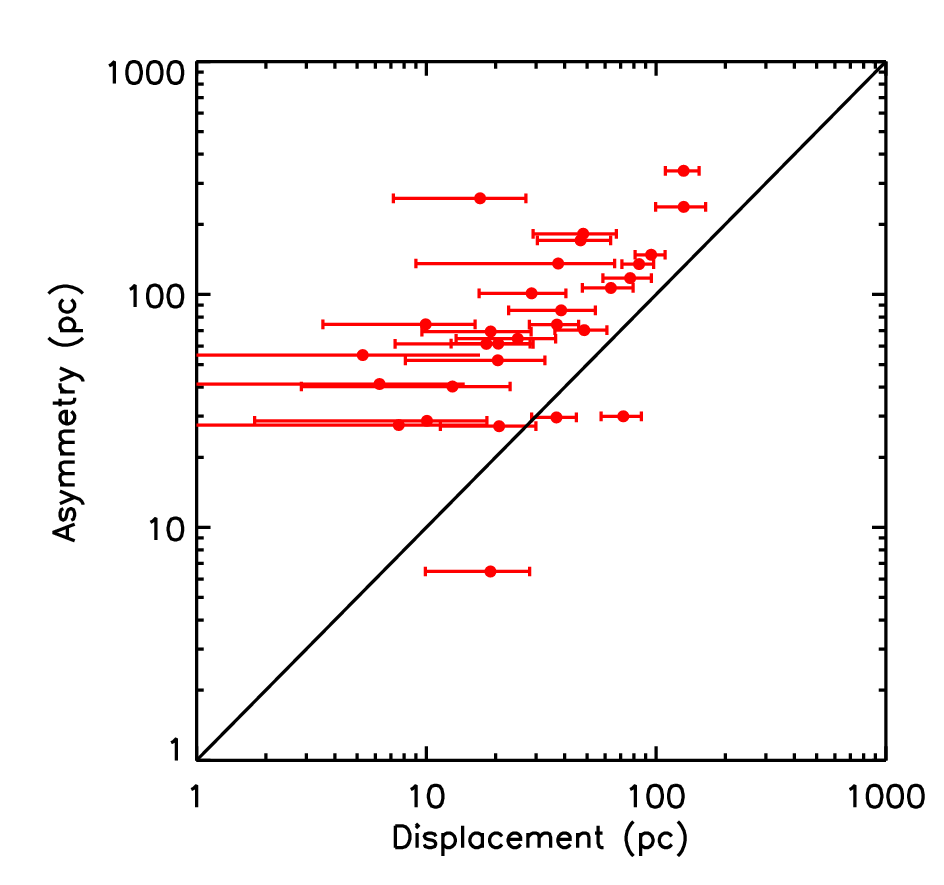}
  \caption{We plot distance vs displacement in the left panel to study systematic
effects due to lower resolution for more distant galaxies.
In the right panel, we plot the
correlation between red giants displacement and degree of asymmetry of the
galaxy (the straight line shows y=x). Note how the asymmetry is bigger than
the displacement in almost all the cases, 
suggesting that the displacement is just a result of internal structure.}
\label{fig:rgsyst}
\end{figure*}

In the left panel of
Figure \ref{fig:rgsyst} we explore whether the observed displacement
between the red giant and main sequence populations
is correlated with the distance to the galaxy.  The data shows a
small positive correlation between distance and displacement.
To check whether this might affect the conclusions of Section \ref{sec:rgb},
we divide the sample in different distance bins and
reanalyze them in individual bins. We find that our conclusions do not change
significantly even if we consider galaxies at similar distances.

In the right panel of Figure \ref{fig:rgsyst}
 we plot the correlation between the
displacement and a measurement of asymmetry. The asymmetry is defined as the
difference in centroids of main sequence stars at two different isocontours of
stellar number density describing inner and outer part of
the galaxy. We see that red giant displacement is strongly correlated with
this asymmetry measurement, indicating that the effect may be related
to the internal structure of the galaxy rather than being due entirely
to a fifth-force effect.  We note, however, that similar morphological
asymmetry may occur in the presence of modified gravity as well.
However, if we find the presence of similar effects in a screened
galaxy sample then we will able to separate the modified gravity effects from
internal structure of the galaxy.


We also checked if the displacement is correlated with the number of stars
(i.e.\ shot-noise), the ratio of red giants compared to the number of main
sequence stars,
the luminosity of the galaxy (indicative of mass and size),  the HI line
width (also indicator of mass), filters used to determine red giants or
magnitude limit in the whole sample which depends on the
camera used. None of these observables show appreciable correlation with
the displacement of red giants.

We examined the effect of differences in
the area that we used to calculate the centroid of main sequence stars and red
giants. Usually the galaxy is not well centered in the image, that is the
reason why we have to study systematics in the definition of 
the galaxy limiting area. Although it is easier to define isocontours in
number counts using all the stars, there are a lot of interlopers (faint stars)
that do not belong to the galaxy. We recalculated the centroid
after gradually removing the faintest stars. We include galaxies only if their
centroids converge after removing 20\% of faintest stars.
We also test if crowded regions can bias the result based on flags defined in
\citet{dalcanton09}. Due to longer exposure times, large number of faint
foreground star will be detected which leads to poor photometry of individual
stars. Moreover, we have calculated the centroid of main sequence stars
weighting by luminosity, and still the displacement of red giants is consistent.
Finally, we change our definition of red giants by going 0.5
magnitudes fainter, and we exclude bright stars from the definition of
main sequence stars, and conclusions remain the same.

We also checked if the displacement is related to the number of stars
(shot-noise), the ratio of red giants compared to the number of main sequence
stars,
the luminosity of the galaxy (indicative of mass and size),  the HI line
width (also indicator of mass), filters used to determine red giants or
magnitude limit in the whole sample which depends on the
camera used. None of these observables are correlated with the
displacement of red giants.

\section{Measurement of Warps}
\label{ap:warp}
In this section we describe the systematics effects related to the measurement of warps in
section \ref{sec:warp}.
We explore the bias due to distance effects, inclination effects,
and environment.  Finally, we propose a third warp estimator
and show that the conclusions of Section \ref{sec:warp} are unchanged.
 
\subsection{Distance Bias}
\label{ap:distance-systematic}
\begin{figure*}
\begin{center}
  \includegraphics[scale=0.37]{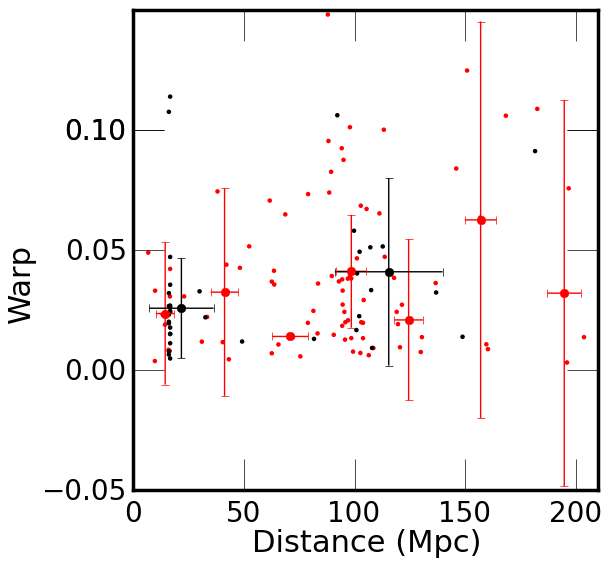}
  \includegraphics[scale=0.37]{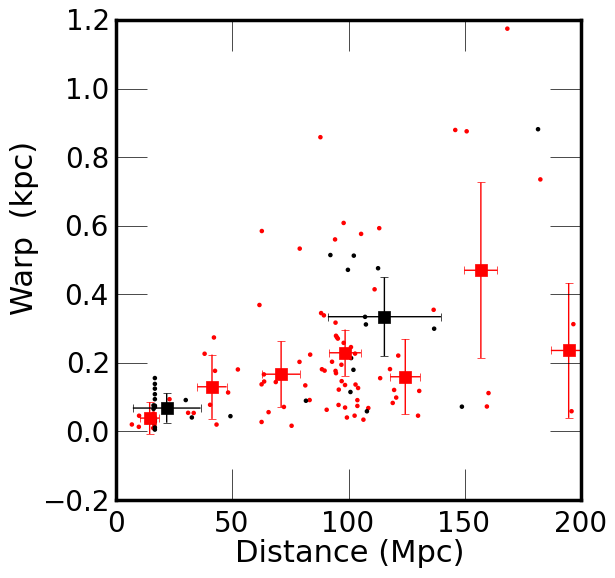}
  \caption{Systematic effects involved in the warp test. The left panel shows the
warp ($w_1$) vs distance (Eq. \ref{warp-param}), right panel shows
warp in kpc vs  distance (Eq. \ref{warp-param2}) for screened and
unscreened galaxies. If the screened sample is systematically
closer than unscreened galaxies, then we measure a lower value of warp for
screened sample compared to unscreened sample, which goes in the same direction
as the modified gravity prediction.}
\label{warp-syst}
\end{center}
\end{figure*}

A number of systematic effects may be involved in the warp analysis.
The first is bias related to distance.
In our sample there exists a strong correlation between
distance and average physical galaxy size:
this is due to the fact that it is difficult to observe physically
smaller galaxies at higher distances. If the measured warp depends on the
size of the galaxy, then the comparison between screened and unscreened
samples should be done as a function of distance to limit systematic effects
due to the different samples.
We check the distance bias by plotting the warp parameter against
distance to the galaxies. The left panel of Figure \ref{warp-syst} shows that
the normalized warp parameter, $w_1$, is independent of the distance to
the galaxy, as expected.

On the other hand, the estimator $w_2$ does depend
on the size of the galaxy: two galaxies with identical $w_1$ curves
will differ in $w_2$,
depending on the size of the galaxy. Therefore, the
size-distance relation implies that the estimated values of $w_2$ will be
higher at larger distances. The right panel of Figure \ref{warp-syst}
shows that this is indeed true. It should be noted that we have a
systematically higher fraction of screened galaxies at smaller distance. We
found that these galaxies belong to the Virgo cluster at distance $\sim 17$
Mpc. Therefore, when we consider all the galaxies irrespective of their size,
the screened sample has a smaller average value of $w_2$ compared to that of
unscreened galaxies due to the large fraction of screened galaxies at smaller
distances. That is, 
the marginal difference between $w_2$ distributions of screened
and unscreened samples can be explained by this effect (compare figure
\ref{fig:warp-distr} and Figure \ref{warp-syst}).

\subsection{Inclination Bias}
\label{ap:inclination-systematic}
\begin{figure}
\begin{center}
  \includegraphics[scale=0.4]{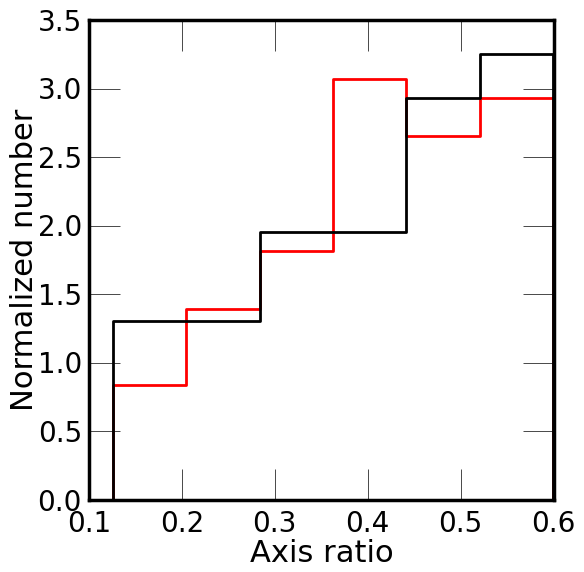}
\caption{The distribution of axis ratio for screened and
unscreened galaxies with $f_{R0}=2\times10^{-7}$. We find no difference  between
the two samples.}
\label{warp-incl}
\end{center}
\end{figure}

Another potential systematic
effect may be related to the inclination of the galaxy.
While the warp can be well-estimated in edge-on galaxies,
the apparent warp may be nearly zero for warped face-on galaxies. This is the
source of the bias introduced by inclination.  If the unscreened galaxies
are systematically more inclined than the screened galaxies, then the warp
parameter estimation will be more robust for unscreened galaxies. 
We therefore, check whether there is any systematic difference between the
distributions of axis ratio for screened and unscreened galaxies. Figure
\ref{warp-incl} shows that the axis ratio
of both screened and unscreened galaxies are distributed similarly. So we
conclude that inclination affects the screened and unscreened galaxies
equally, and therefore is unlikely to bias our result.

\subsection{Environmental Bias}
An additional test we can perform with the data is to check whether there is
any noticeable difference in the fraction of galaxies displaying
U-shaped warps between the screened and unscreened samples.
In the modified gravity scenario we expect
that the fraction of U-shaped warps will be larger in the unscreened sample
than in the screened sample. For this test we classify galaxies as U-shaped
only if the galaxy has more than 3\% warp, i.e. if
$w_1 > 0.03$. As above, if the signs of left and right side warp
are different then we classify them as U-type.

\begin{figure}
\begin{center}
\includegraphics[scale=0.35]{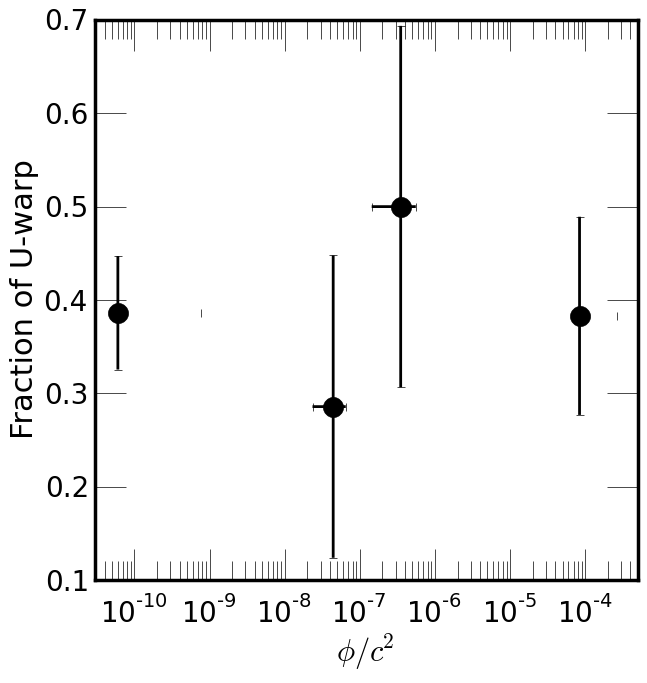}
\caption{The fraction of galaxies displaying U-shaped warps
  vs the magnitude of the external potential. In this plot, galaxies are classified as U-shaped only
  if the warp is greater than or equal to 3\%.}
\label{fig:warp-stren-frac}
\end{center}
\end{figure}

Figure \ref{fig:warp-stren-frac} shows that the fraction of U-shaped
warps does not depend on the screening level. This implies that any
physical process which causes the formation of
U-shaped warp works irrespective of the local
density of the galaxy. It should be noted that this result does not vary with
the choice of lower threshold value of the warp (i.e. $w_1 > 3$\%) to define
the U-warp.  In the following section, we further show that the fraction
of S-shaped warps is also consistent between the screened and unscreened
sample.

\subsection{Distribution of warps}
\label{ap:warp-distri}
\begin{figure*}
\begin{center}
  \includegraphics[scale=0.34]{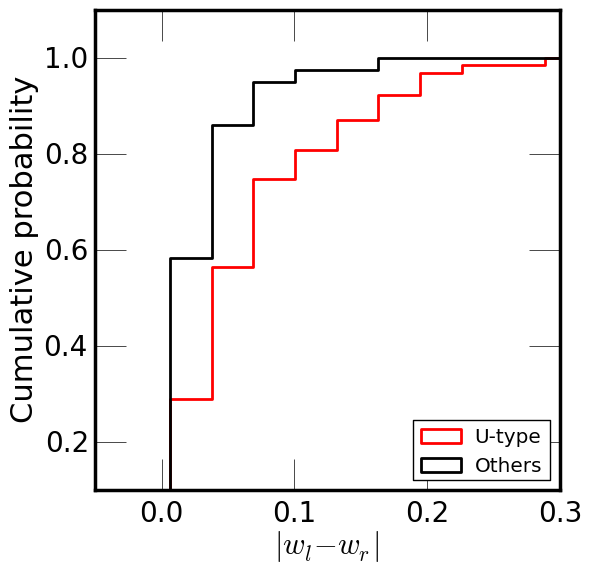}
\includegraphics[scale=0.35]{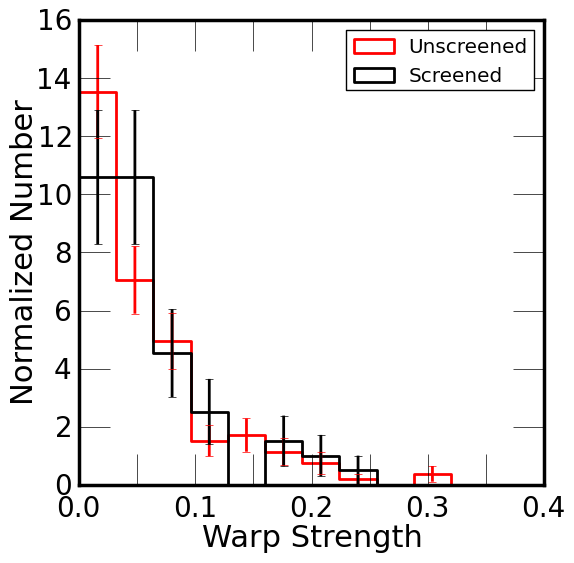}
\includegraphics[scale=0.35]{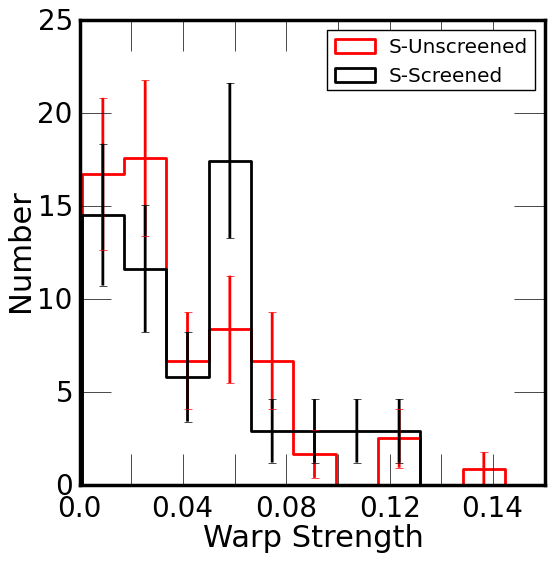}
\caption{Left : The cumulative histrogram of $w_e$ (see text for the
definition). The U-type warps can be distinguished  with $w_e > 0.1$. Middle
panel shows the distributions of $w_e$ in screened and unscreened regions for
$f_{R0}=2\times10^{-7}$. The right panel shows the distributions of S-type
warps in screened and unscreened regions with $f_{R0}=2\times10^{-7}$.}
\label{warp-cumu-1}
\end{center}
\end{figure*}

In Section \ref{sec:warp} we have shown that the distribution of
U-type warp does not depend on the the environment. However, we include
only U-type warp in that analysis. Here we check whether the conclusion in
 section \ref{sec:warp} still holds if we include all kind of
warp in the analysis. For this purpose we define the \textit{effective}
total warp ($w_e$) of galaxies as the absolute value of the difference
between the warp of left and right hand sides of the galaxy. As we
explained before the U-type galaxies will have opposite signs for left and 
right side waprnesses and S-type/N-type will have same signs for both sides.
Therefore, our definition of $w_e$ will leads to larger value for a symmetric
U-type warp and zero for other cases. However, the other type of warps can also
give smaller non-zero values for $w_e$ depending on the asymmetry of the warp
curve. The same applies to U-type warp where the $w_e$ can be smaller for weak
warp curves. We find that almost all galaxies, except U-type warps, have
$w_e < 0.1$ with warp definition Eq \ref{warp-param}. It is shown in the
left panel of Figure \ref{warp-cumu-1}. Also, we found that the distribution of
$w_e$ does not have any dependence on the environment and it is shown in the
middle panel of Figure \ref{warp-cumu-1}. We also found that there is no 
noticeable difference between the distributions of S-type warps in screened and
unscreened regions and it is shown in the right panel of the figure. We found
similar results if we use Eq \ref{warp-param2} to define
warp. In that case $w_e = 0.5$ kpc separate U-type warps from the rest. 

\begin{figure}
\begin{center}
  \includegraphics[scale=0.4]{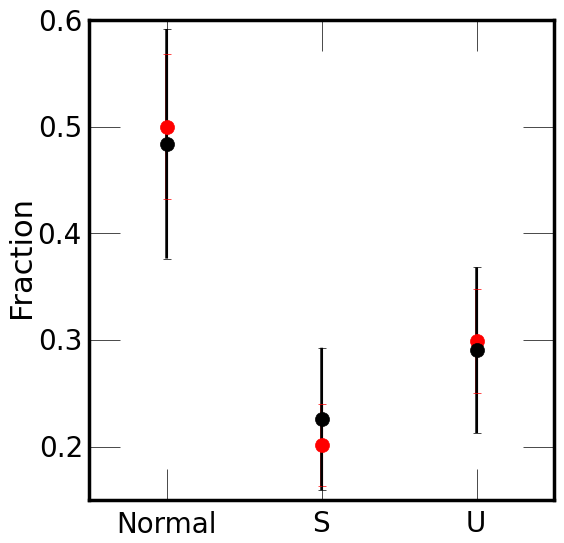}
 \caption{The fraction of normal, S-type and U-type galaxies in screened and
unscreened regions. We find no noticeable difference between the fraction of
different types in screened and unscreened regions. The error in the plots are
calculated by assuming Poissonian noise on the number of galaxies.}
\label{warp-frac}
\end{center}
\end{figure}

In Figure \ref{fig:warp-stren-frac} we show the fraction of galaxies
with U-type warps in different environments. We check whether the fraction of
normal galaxies and fraction of S-type galaxies have any dependence on their
environment and the result is shown in Figure \ref{warp-frac}. We classify
a galaxy as normal if the warp is less than 5\%.  We find that the
fractions of different types of galaxies is independent of their
environment. The result is insensitive to
the definition of normal galaxy.

\begin{table*}
\begin{center}
\begin{tabular}{ccccccccc}
\textbf{Type} & $f_{R0}$ & \multicolumn{2}{|c|}{\textbf{Number}} &
\multicolumn{2}{|c|}{\textbf{$\langle w_1\rangle$}} &
\textbf{$\hat{w}_{MG}$} & \textbf{Prediction} \\
 & & $N_S$ & $N_U$ & $w_S$ & $w_U$ & & $\Delta G/G=1/3$,$\Delta G/G=1$ & \\
\hline
All    & 2e-07 & 59 & 158 & $0.051 \pm 0.023$ & $ 0.050 \pm 0.014$ & $-0.001 \pm
0.027$ & \\
All    & 1e-06 & 143 & 352 & $0.038 \pm 0.014$ & $ 0.046 \pm 0.009$ & $0.008 \pm
0.017$ & \\
U & 2e-07 & 48 & 78 & $0.031 \pm 0.018$ & $ 0.036 \pm 0.015$ & $0.006
\pm 0.023$ & $0.003_{0.001}^{0.003}$, $0.006_{0.003}^{0.006}$\\
U & 1e-06 & 89 & 229 & $0.024 \pm 0.012$ & $ 0.029 \pm 0.009$ & $0.005
\pm 0.015$ & $0.002_{0.001}^{0.002}$, $0.004_{0.002}^{0.004}$ \\

\hline
\\
\\
 & & & & \multicolumn{2}{|c|}{\textbf{$\langle w_2\rangle$}~~(in kpc)} & &  \\
 & & & & $w_S$ & $w_U$ & & & \\
\hline
U $h_r=0.5$ kpc & 2e-07 & 16 & 11 & $0.048 \pm 0.036$ & $0.055 \pm 0.051$ &
$0.007 \pm 0.062$ & $0.00 \pm 0.00$, $0.00 \pm 0.00$ \\
U $h_r=1.5$ kpc & 2e-07 & 14 & 35 & $0.151 \pm 0.098$ & $0.129 \pm 0.047$ &
$-0.022 \pm 0.109$ & $0.02 \pm 0.02$, $0.03 \pm 0.04$ \\
U $h_r=2.5$ kpc & 2e-07 & 11 & 24 & $0.182 \pm 0.091$ & $0.330 \pm 0.104$ &
$0.148 \pm 0.138$ & $0.03 \pm 0.02$, $0.07 \pm 0.04$ \\
U $h_r=3.5$ kpc & 2e-07 & 6 & 6 & $0.312 \pm 0.000$ & $0.288 \pm 0.177$ &
$-0.024 \pm 0.177$ & $0.08 \pm 0.06$, $0.17 \pm 0.13$ \\

U $h_r=0.5$ kpc & 1e-06 & 16 & 13 & $0.046 \pm 0.035$ & $0.053 \pm 0.041$ &
$0.006 \pm 0.054$ & $0.00 \pm 0.00$, $0.00 \pm 0.00$ \\
U $h_r=1.5$ kpc & 1e-06 & 16 & 60 & $0.094 \pm 0.057$ & $0.144 \pm 0.043$ &
$0.050 \pm 0.071$ & $0.01 \pm 0.02$, $0.03 \pm 0.03$ \\
U $h_r=2.5$ kpc & 1e-06 & 15 & 73 & $0.145 \pm 0.074$ & $0.215 \pm 0.047$ &
$0.070 \pm 0.087$ & $0.02 \pm 0.02$, $0.04 \pm 0.03$ \\
U $h_r=3.5$ kpc & 1e-06 & 12 & 50 & $0.255 \pm 0.092$ & $0.279 \pm 0.066$ &
$0.024 \pm 0.113$ & $0.04 \pm 0.05$, $0.09 \pm 0.09$ \\
\end{tabular}
\end{center} 
\caption{Summary of warp tests. The columns are: 1. {\bf Type:} the 
type of warp included in the analysis. All - included all galaxies, U - used
only U-type warps. 
2. $f_{R0}$ - the value of $f_{R0}$ parameter used to classify
screened and unscreened galaxies 
3. {\bf Number:} - $N_S$, $N_U$ are the number of
galaxies in screened and unscreened regions. 
4. $w_S$, $w_U$ are the average
value of the two  $w_1$ or $w_2$ in screened and unscreened galaxies 
5. $\hat{w}_{MG}$ - estimated warp due to modified gravity. 
6. {\bf Prediction:}  the theoretical predictions for  unscreened galaxies 
with $\Delta G/G=1/3$
and $\Delta G/G=1$.
The upper part of the table is for the dimensionless warp parameter $w_1$ 
(defined in Eqn. \ref{warp-param}) and lower part is for the parameter $w_2$
(defined in Eqn. \ref{warp-param2}). In the second part of the table we group
U-type galaxies based on their half light radius ($h_r$) to compare the
warp. See text for details. The estimated value of $\hat{w}_{MG}$ is
consistent with GR in all cases, but the error bars are too large to detect the
MG predictions shown in the last column. 
}
\label{tab:warp}
\end{table*}

\section{Examples of interesting observations}
\label{ap:curves}
\begin{figure*}
\centering{
\includegraphics[trim= 0cm 0cm 0cm 0cm, clip = true, width=0.3\textwidth]
{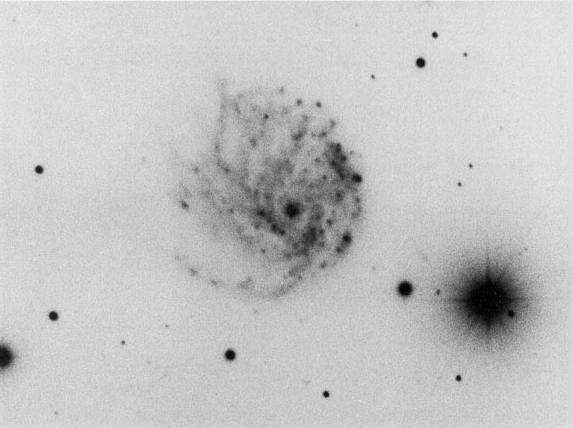}}
\centering{
\includegraphics[trim= 0cm 0cm 0cm 0cm, clip = true, width=0.3\textwidth]
{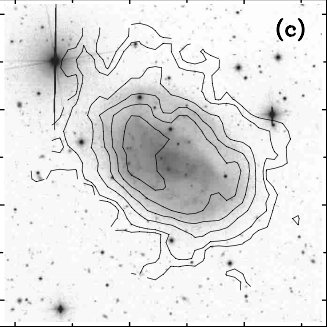}}
\caption{{\it Left panel:} Optical image of dwarf galaxy UGC 3740 which is
unscreened for $f_{R0}=10^{-6}$. The galaxy is asymmetric in the optical. The \ha~
rotation curve also shows strong asymmetry. 
{\it Right panel:} Comparison of HI and optical images of UGC 4325. The HI
image shows some asymmetry. However, this galaxy does not have any nearby neighbors to which we could attribute an external force. The images are taken from \citet{figgs2008,swaters2002b}
\label{fig:eg-image-1}}
\end{figure*}

\begin{figure*}
\centering{
\includegraphics[trim= 0cm 0cm 0cm 0cm, clip = true, width=0.3\textwidth]
{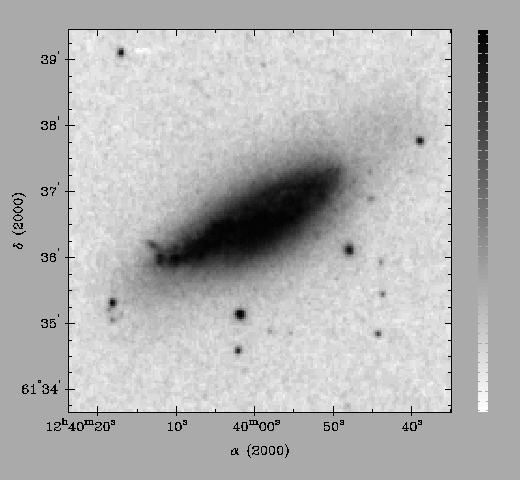}}
\centering{
\includegraphics[trim= 0cm 0cm 0cm 0cm, clip = true, width=0.3\textwidth]
{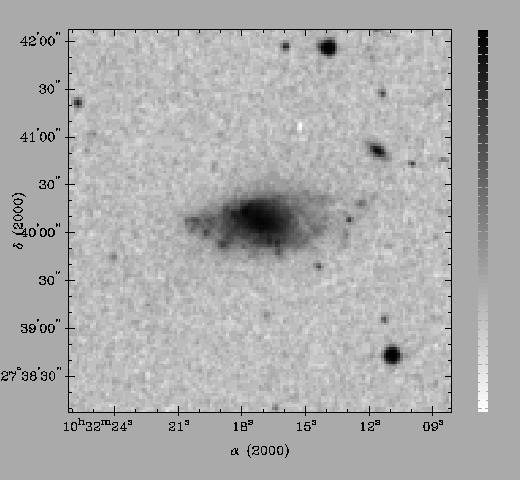}}
\centering{
\includegraphics[trim= 0cm 0cm 0cm 0cm, clip = true, width=0.3\textwidth]
{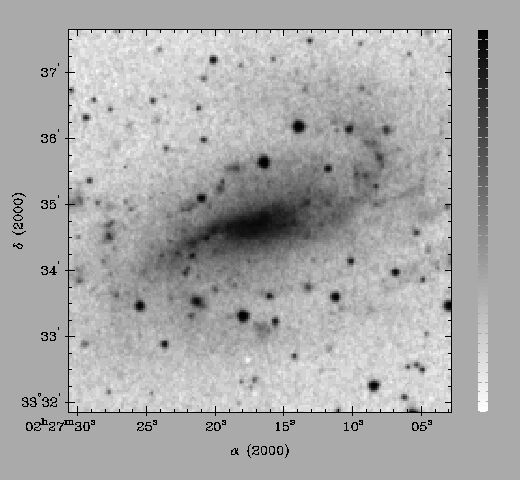}}
\caption{Optical images of  galaxies with the most asymmetric \ha~
velocity curves: {\it Left panel:} 
UGC 7831 which is unscreened for $f_{R0}=10^{-6}$ and shows some warp.  
{\it Middle panel:} UGC 5721. {\it Right panel:} UGC 1913. 
The images are taken from \citet{figgs2008}
\label{fig:eg-image-2}}
\end{figure*}

\begin{figure*}
\centering{
\includegraphics[trim= 0cm 0cm 0cm 0cm, clip = true, width=0.9\textwidth]
{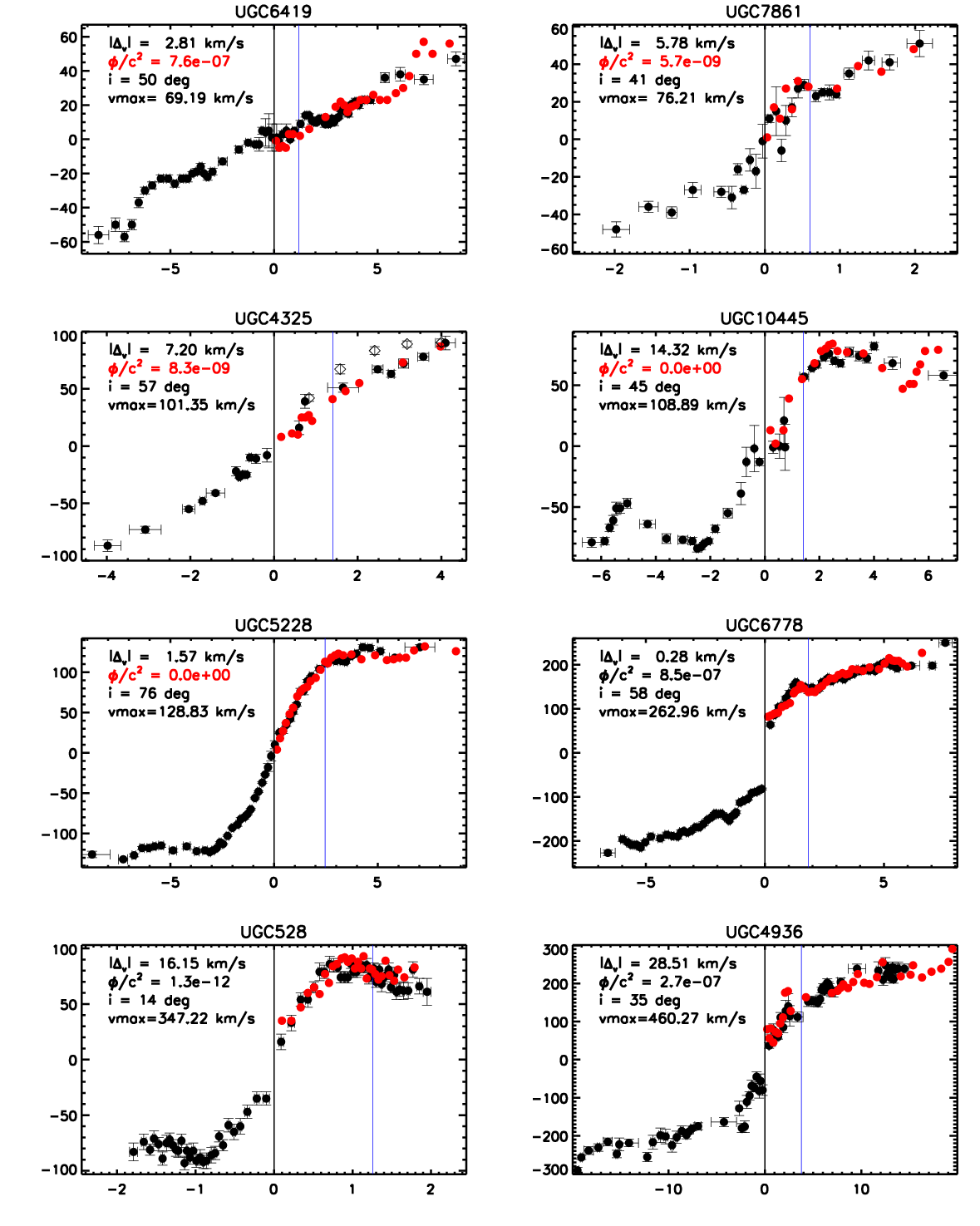}}
\caption{Examples of \ha~ symmetric rotation curves for GHASP galaxies.  The x-axis  is in kpc, and the y-axis shows the measure rotation velocity in km/s (before 
correcting for inclination). The
left side points of the rotation curves are shown as red points for comparison 
to the right side. We indicate the
name of the galaxy,  $\Delta v$ defined in the text and the external 
potential due to neighbors within 3 Mpc. 
The potential is  
in red if the galaxy is unscreened, 
and black if it is environmentally screened, for $f_{R0}=10^{-6}$. 
We label the inclination and the averaged maximum velocity of the galaxy
once corrected for inclination.
Diamonds show HI curves, when available. The blue vertical lines show the range 
used to calculate $\Delta v$, based on its half-light radius. Note that for UGC 4325
the velocity in HI is larger than in \ha{}.  
\label{fig:GHASPsym}}
\end{figure*}

\begin{figure*}
\centering{
\includegraphics[trim= 0cm 0cm 0cm 0cm, clip = true, width=0.9\textwidth]
{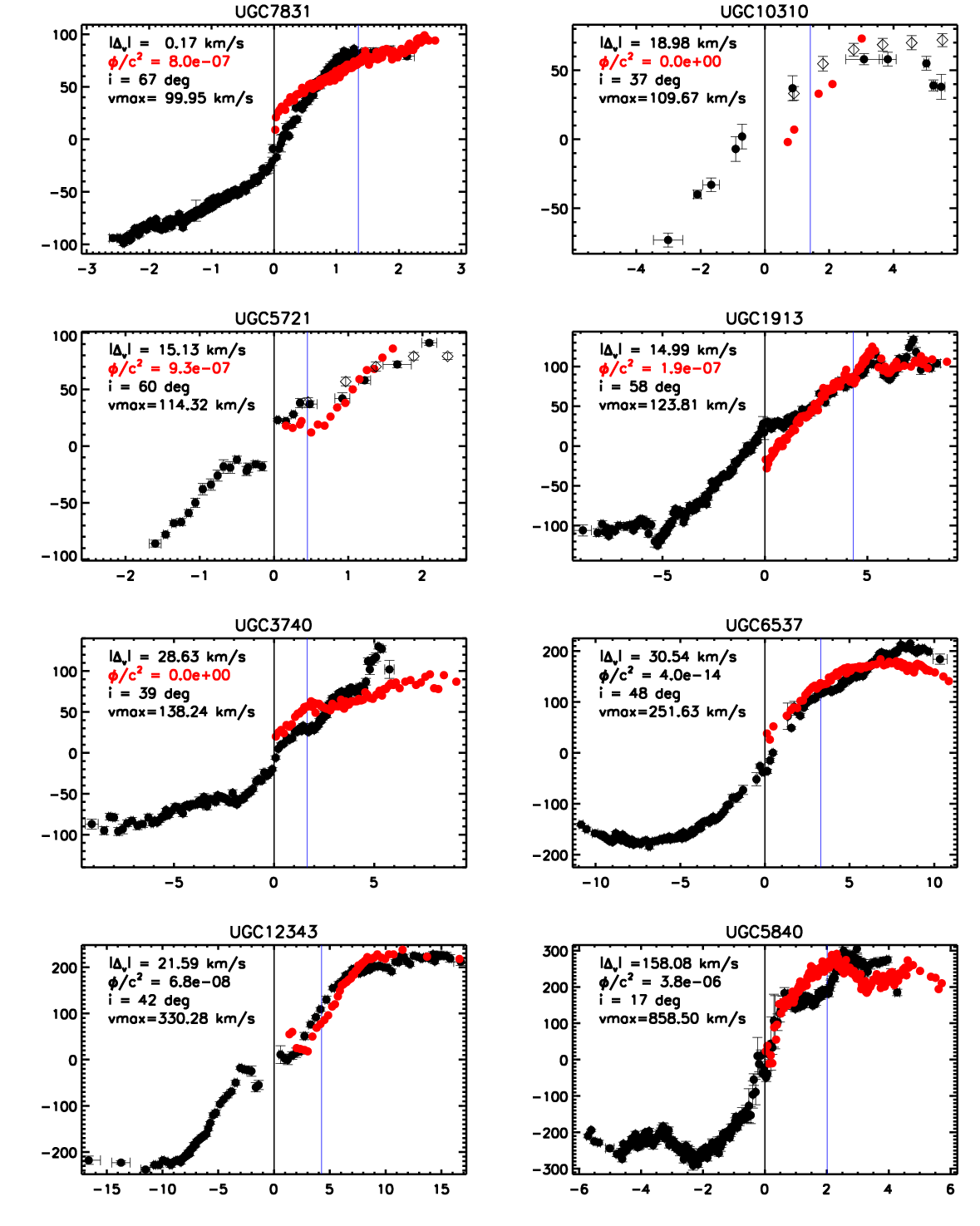}}
\caption{Same as Figure \ref{fig:GHASPsym} for smoothly asymmetric curves, 
selected visually. These slow varying asymmetries could be created by modified
gravity.  
\label{fig:GHASPasym}}
\end{figure*}

We present some examples of optical images (Figures \ref{fig:eg-image-1} \&
\ref{fig:eg-image-2}) and rotation curves (Figures \ref{fig:GHASPsym} \&
\ref{fig:GHASPasym}) of unscreened galaxies from the literature which look
interesting in the context of modified gravity.

\end{document}